\documentclass[11pt,a4paper]{article}
\usepackage{amssymb,amsfonts,amsmath,amsthm}
\usepackage[latin1]{inputenc}
\usepackage[english]{babel}
\usepackage[left=2.7cm,right=2.7cm,bottom=2.67cm,top=2.67cm]{geometry}
\usepackage{dsfont}
\usepackage{mathrsfs}
\usepackage{natbib}
\usepackage{adjustbox}
\usepackage{color}
\usepackage{sectsty}
\usepackage[implicit=false]{hyperref}
\usepackage[normalem]{ulem}
\RequirePackage[OT1]{fontenc}
\usepackage{graphicx}
\usepackage{multirow}
\usepackage{subfigure}
\usepackage{graphicx}
\usepackage{placeins}
\usepackage{enumitem}
\usepackage{mathrsfs}
\usepackage[table]{xcolor}

\theoremstyle{plain}

\theoremstyle{definition}

\newcommand{\R}{\mathds R}
\newcommand{\N}{\mathds N}

\newcommand{\F}{\mathscr{F}}

\newcommand{\bs}[1]{\boldsymbol{#1}}

\long\def\sfootnote[#1]#2{\begingroup%
\def\thefootnote{\fnsymbol{footnote}}\footnote[#1]{#2}\endgroup}

\def\bfootnote{\xdef\@thefnmark{}\@footnotetext}

\begin{document}
\pagestyle{myheadings} 

\thispagestyle{empty}
{\centering
\Large{\bf Mitigating the choice of the duration in DDMS models through a parametric link }\vspace{.5cm}\\ 
\normalsize{ {\bf Fernando Henrique de Paula e Silva Mendes${}^{\mathrm{a,}}$\sfootnote[1]{Corresponding author. This Version: \today}\let\thefootnote\relax\footnote{\hskip-.15cm${}^\mathrm{a}$Graduate Program in Statistics - Federal University of Rio Grande do Sul.}, Douglas Eduardo Turatti${}^{\mathrm{b}}$\let\thefootnote\relax\footnote{\hskip-.15cm${}^\mathrm{b}$Aalborg University Business School - Aalborg University.} and Guilherme Pumi${}^{\mathrm{a}}$
}}
 \\
\let\thefootnote\relax\footnote{E-mails: mendes.fernandohenrique@hotmail.com (F.H.P.S Mendes), guilherme.pumi@ufrgs.br (G. Pumi) and det@business.aau.dk (D.E. Turatti)}}\\
\vskip.3cm
\begin{abstract}
	
One of the most important hyper-parameters in duration-dependent Markov-switching (DDMS) models is the duration of the hidden states. Because there is currently no procedure for estimating this duration or testing whether a given duration is appropriate for a given data set, an ad hoc duration choice must be heuristically justified. In this paper, we propose and examine a methodology that mitigates the choice of duration in DDMS models when forecasting is the goal. Two Monte Carlo simulations, based on classical applications of DDMS models, are employed to evaluate the methodology. In addition, an empirical investigation is carried out to forecast the volatility of the S\&P 500, which showcases the capabilities of the proposed model.

\vspace{.2cm}

\vspace{.2cm}
\end{abstract}
\section{Introduction}
Many authors have suggested variants of the Markov-switching model since the seminal paper of \cite{hamilton_new_1989}. In this context, \cite{durland_duration-dependent_1994} proposed the duration-dependent Markov-switching (DDMS) model based on a higher-order Markov chain that allowed state transition probabilities to be duration-dependent. Initially applied to investigate business cycle, such as if the continuation of expansions or recessions is dependent on how long the economy has been in those regimes, this modeling approach also includes bull and bear market stock identification \citep{maheu_identifying_2000} and foreign exchange volatility estimation \citep{maheuvolatility2000} since duration can be included as a conditioning variable in the conditional mean/variance equations. Based on that, the empirical literature has also explored generalizations of the DDMS model, broadening the baseline duration dependence structure in different directions \citep[see, for example,][among others.] {Lam2004, pegalatti,bejaoui_revisiting_2016}.

One of the key components in DDMS models is the duration value to be used in the application, which is a user-specified parameter that cannot be estimated. The duration selection in the existing literature is either arbitrary, under the argument that the impact of the duration dependence vanishes after selecting a ``reasonably'' large value or based on a grid search aimed at maximizing the likelihood. The difficulty in selecting and justifying the duration is the main criticism of DDMS models, often hindering their use in applications. In contrast to the existing literature, \cite{jp10} showed, using a Bayesian approach, that duration structures are not necessarily monotonic and, therefore, cannot be described by the conventional models. The authors restrict their approach to a bull and bear market analysis in the same spirit of \cite{maheu_identifying_2000} through a country-by-country stock data analysis; however, they do not extend their approach to other DDMS-type models and  restrict the study to an in-sample design.

In this context, the main contribution of the present paper is to develop and examine a strategy for estimating DDMS models that attenuates the problem of duration selection and justification by allowing for a completely arbitrary choice. Our approach is based on  the use of the asymmetric Aranda-Ordaz parametric link function \citep{aranda} instead of the (fixed) logit link in the transition probabilities in DDMS models. The idea behind this approach is that any potentially incorrect duration choice is compensated for by the parameter in the link, increasing model flexibility by letting the data ``tell'' which is the ``best'' link, allowing the model to capture the latency of the Markov chain endogenously, improving the fit and forecasting accuracy under misspecification of the duration.

To illustrate our approach empirically, we extend the research that evaluates alternative volatility modeling and forecasting methods for S\&P500 daily log returns by broadening the traditional DDMS specifications to include the Aranda-Ordaz link. More precisely, we conduct a pairwise Diebold-Mariano-West statistical test for the one-day ahead forecast stock volatility from April 2018 to January 2020, comprising 443 out-of-sample observations. For robustness checking, we apply different volatility proxies and loss functions commonly found in the literature. Overall, our modeling approach outperforms the benchmark logit specification under certain reasonable conditions, mitigating the uncertainty caused by the duration choice. In addition, we also compare the proposed specification to Garch-type models. We did not find statistical evidence to reject the null hypotheses of equal predictive accuracy, while these results were not observed for the models with fixed logit link.

To evaluate the performance of the proposed approach, we conduct a Monte Carlo simulation study inspired by two classical applications of DDMS models: (i) a bull and bear market identification for the in-sample stock market cycles probabilities and (ii) a point volatility forecasting exercise conducted for different out-of-sample horizons. In general, we observe that the proposed Aranda-Ordaz approach improves estimates in different directions; however, a typical result is the advantage (in proportion) of the Aranda-Ordaz over the logit for higher likelihood values. Over the stock market probabilities, we observe that the Aranda-Ordaz model also presents superior frequency with probability closer to the true values under duration misspecification. From the forecasting perspective, the application of the Aranda-Ordaz is advantageous regardless of the true value of the duration. This indicates that even when the correct duration is used, the model based on the logit cannot provide the ``best possible'' forecasts.

The remainder of this paper is organized as follows: Section 2 presents the proposed model; Section 3 describes the optimization algorithm designed to perform the maximum likelihood estimation. in Section 4 we present a Monte Carlo simulation study whereas in Section 5 we present an empirical application of the proposed model to daily returns of the S\&P 500 index. Section 6 concludes the paper.

\section{The proposed model}
Based on \cite{maheu_identifying_2000, maheuvolatility2000}, we start by considering a simple stochastic volatility model given by
\begin{equation*}
Y_t = \mu ({S_t}) + \sigma\big(S_t,D(S_t)\big)Z_t,
\end{equation*}
where $S_t$ denotes the state mixing variable, $D(S_t)$ is the duration of the state $S_t$, at time $t$, and $Z_t\sim N(0,1)$ is an i.i.d. error term. The duration $D(S_t)$ depicts the length of a run of realizations of a particular state and, in principle, could grow very large causing estimation problems and numerical instability. To avoid such problems, we define
\begin{equation*}
D(S_{t}):=\min\big\{D(S_{t-1})I(S_t=S_{t-1})+1,\tau\},
\end{equation*}
where $\tau\in\N$ is a user chosen threshold such that the duration is accounted for up to time $\tau$, and $I$ is the indicator function. The transition probabilities associated to the latent states $S_t$ are parameterized using a similar approach as in generalized linear model by means of a link. The most commonly applied link is the logit, which yields
\begin{equation}\label{transprob}
P\big(S_{t}=i|S_{t-1}=i, D(S_{t-1})=d\big) = \frac{\exp \left(\gamma_{1}(i)+\gamma_{2}(i) (d\wedge\tau)\right)}{1+\exp \left(\gamma_1(i)+\gamma_{2}(i) (d\wedge\tau)\right)}, \quad i=0,1,
\end{equation}
where $d\wedge\tau = \min\{d,\tau\}$ and $\gamma_{j}(i)$, $i,j\in\{1,2\}$, are parameters to be estimated.

In this work we propose to parameterize the transition probabilities upon applying a twice differentiable one-to-one parametric link function $g(\cdot,\lambda):(0,1)\rightarrow \R$, in the same generalized linear model approach as before. That is, we consider
\begin{equation*}
g\big(P\big(S_{t}=i | S_{t-1}=i, D(S_{t-1})=d\big);\lambda\big) = \gamma_{1}(i)+\gamma_{2}(i) (d\wedge\tau), \quad i=0,1,
\end{equation*}
or, equivalently,
\begin{equation}\label{transAO}
P\big(S_{t}=i | S_{t-1}=i, D(S_{t-1})=d\big) = g^{-1}\big(\gamma_{1}(i)+\gamma_{2}(i) (d\wedge\tau);\lambda\big) \quad i=0,1,
\end{equation}
where $\lambda$ is a parameter to be estimated from the data. One of the most commonly applied parametric link function is the so-called asymmetric Aranda-Ordaz link function \citep{aranda}, given by
\begin{equation*}
g(y;\lambda)=\log\bigg(\frac{(1-y)^{-\lambda}-1}{\lambda} \bigg),
\end{equation*}
for $y\in(0,1)$ and $\lambda>0$, whose inverse is given by
\begin{equation*}
g^{-1}(x;\lambda) = 1-\big( 1+\lambda e^x\big) ^{-\frac{1}{\lambda}},
\end{equation*}
for $x\in\R$. Observe that $\displaystyle{\lim_{\lambda\rightarrow 0+}}g^{-1}(x;\lambda)=1-e^{-e^x}$ and $\displaystyle{\lim_{\lambda\rightarrow 0+}}g(x;\lambda)=\log\big(-\log(1-x)\big)$ which is the so-called cloglog link function.

\begin{figure}[ht]	
\centering
\subfigure[$d=3$, $\lambda\in\{1,12\}$ and $i=0$ ]{\includegraphics[width=.45\textwidth]{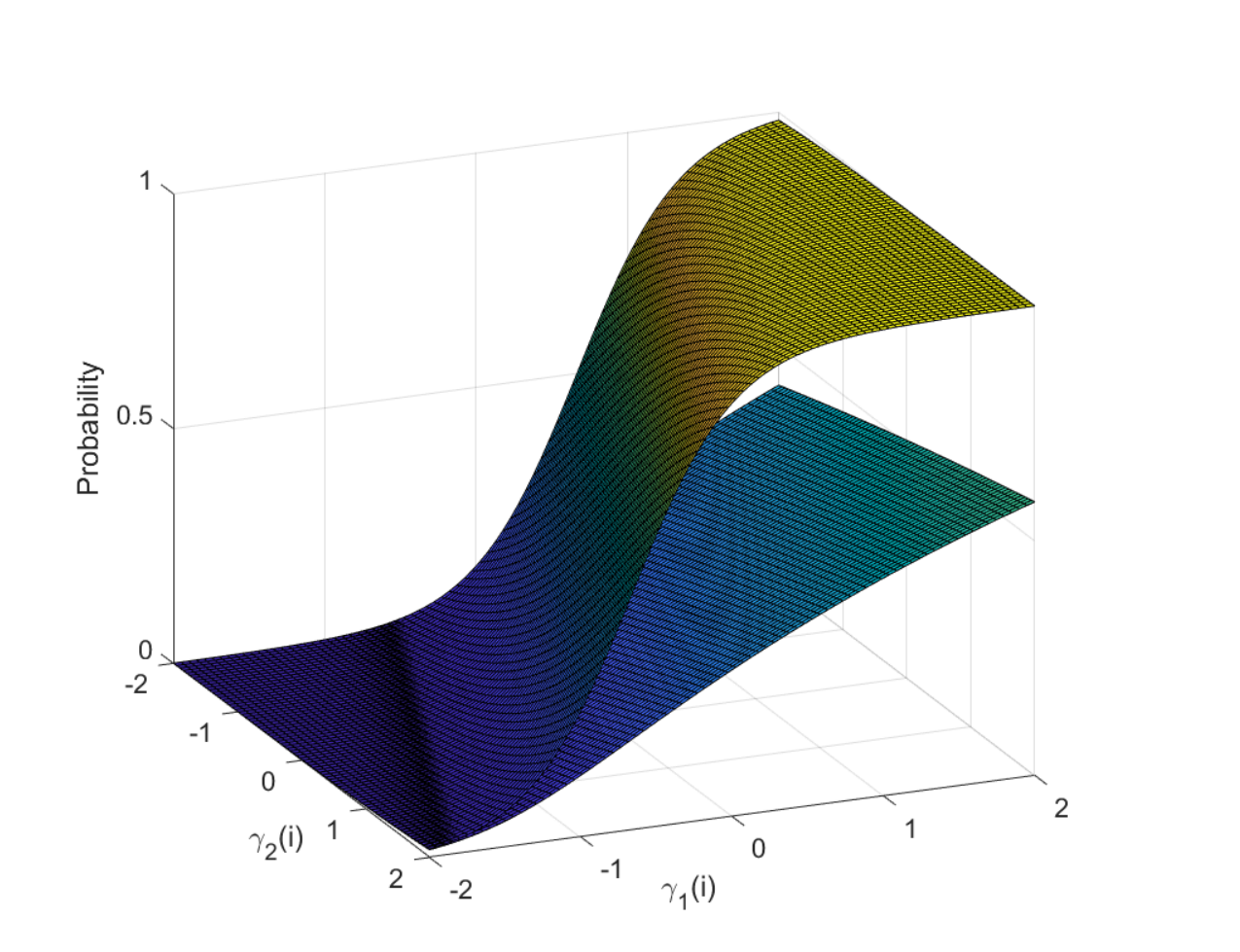}\label{fig1_a}}
\subfigure[$\lambda=1$, $d\in\{3,12\}$ and $i=0$ ]{\includegraphics[width=.45\textwidth]{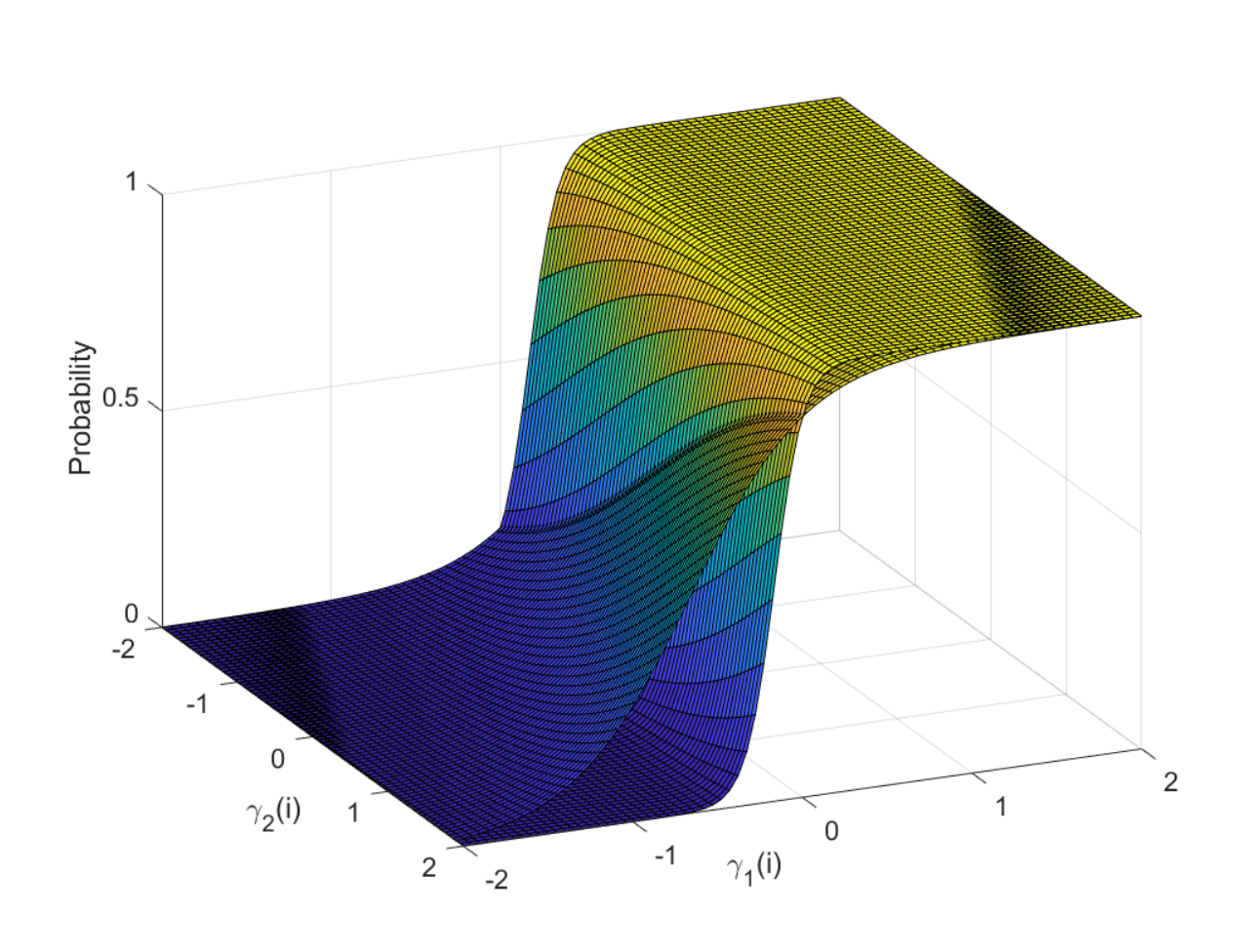}\label{fig1_b}}
\caption{Transition probabilities shape for different values of $\lambda$ and $d$.}\label{fig1}
\end{figure}

To exemplify the effects of the parameters $d$ and $\lambda$ in the transition probabilities \eqref{transAO} as a function of $\gamma_1(0)$ and $\gamma_2(0)$. Figure \ref{fig1_a} presents the case  $d=3$, and $\lambda\in\{1,12\}$ and in Figure \ref{fig1_b}, we have $\lambda=1$ and $d\in\{3,12\}$. In both cases we have transition probabilities parameters $\gamma_j(0)\in[-2,2]$, for $j\in\{1,2\}$.

\FloatBarrier
\section{Optimization Strategy}

Parameter estimation of DDMS models can be challenging and prone to numerical issues. The likelihood function may have multiple local maxima, flat and spiky regions making numerical optimization difficult. Adding an extra parameter through the Aranda-Ordaz link may further complicate the estimation process, given that both the transition parameters and the link function need to be estimated. This suggests that the likelihood function may have issues identifying the transition matrix and the link function parameters, implying it can be near flat on some directions. Additionally, the estimation process in DDMS models with a large duration parameter can also pose computational difficulties. A large duration parameter often leads to a large sparse transition matrix in the extended representation of the DDMS model.  This results in a sparse transition matrix that may approach singularity for several combinations of parameters. Singularity means that some states are given probabilities numerically close to zero, causing the unconditional probabilities to not exist. This results in the likelihood function being undefined at multiple points in the parameter space.

To address these challenges, we developed an optimization algorithm especially tailored to maximize the log-likelihood function of DDMS models. The proposed approach applies a combination of random and grid search techniques for initial parameter values and constrained numerical optimization within defined bounds to tackle multimodality. We also impose nonlinear restrictions to ensure the invertibility of the transition matrix. The details of the algorithm are outlined below.

\begin{enumerate}
\item Find the starting values using a combination of random and grid search: Let $\kappa$ be the number of parameters in the DDMS model. Define a vector $\bs b$ consisting of 100 evenly spaced values between $0.1$ and $10$ for the parameter $\lambda$. For the remaining parameters, create a $(\kappa-1) \times 100$ matrix $C$ with random numbers, where each row is drawn from a continuous uniform distribution. The bounds of the uniform distributions are heuristically defined and  may vary depending on the particular model and dataset.  The resulting draws can then be represented as $\big[ C' \ \ \bs b_{i} \otimes \bs 1_{100}\big]'$, for all $i = 1, \dots, 100$, where $\otimes$ denotes the Hadamard (elementwise) product and $\bs 1_{k}\in\R^k$ denotes a vector of ones.
\item Evaluate the likelihood function at each of the points defined by $\big[ C'\ \ \bs b_{i} \otimes \bs 1_{100}\big]'$, for all $i = 1, \cdots, 100$. Sort the results in decreasing order and store the top $s$ values. Then proceed with the first set of parameters.
\item Let $\bs\theta_0:=(\theta_{0,1},\cdots,\theta_{0,\kappa})'$ be the current vector of starting values for optimization. We look for a local maxima in the domain $[\theta_{0,1} - r, \theta_{0,1} + r]\times\cdots\times [\theta_{0,\kappa} - r, \theta_{0,\kappa} + r]$, where $r$ can either depend on the values of $\bs\theta_0$ or be fixed exogenously. For simplicity, we set $r = r_1$. However, it is important to note that some parameters may have restricted parameter spaces, such as $\lambda$, which must be positive. In such cases, the bounds must satisfy the restrictions on the parameter space.
\item To  guarantee the existence of unconditional probabilities, which are defined by $\pi = (A'A)^{-1} A'\begin{bmatrix}
    \bs 0_{N}\\1\end{bmatrix}$, where $A=\begin{bmatrix}I_{N}-\mathcal P\\ \bs 1_N'\end{bmatrix}$, with $\mathcal P$ denoting the transition matrix for the extended states\footnote{For further details, see Appendix.}, and $\bs 0_N$ the null vector in $\R^N$, ensure that the reciprocal condition number of the matrix $(A'A)$ is above the machine precision. In practice, a small value such as $10^{-9}$ is sufficient to virtually eliminate numerical issues caused by a near singular transition matrix.
\item Use a derivative based numerical optimization method to find the maximum of a nonlinear, multivariate function with bounds, as specified in 3 and nonlinear constraints as defined in 4.
\item The optimization in step 5 is considered successful if the first-order optimality measure is close to zero and the proposed solution does not approach the boundary defined in step 3. We evaluate the proximity of the proposed solution to the bounds by computing the absolute percentage difference. This percentage difference should be above a specified threshold. More specifically, let
    \begin{equation*}
    \ell^-_i:=\frac{|\theta_{1,i} - (\theta_{0,i} - r)|}{|\theta_{1,i}|},\quad\mbox{ and }\quad\ell^+_i:=\frac{|\theta_{1,i}-(\theta_{0,i} + r)|}{|\theta_{1,i}|},
    \end{equation*}
    for $i\in\{1,\cdots,\kappa\}$, where $\theta_{1,i}$ is the proposed value for the $i$th parameter. We say that the proximity criteria is met for a given threshold $\delta>0$ if $\min\{\ell^-_1,\cdots,\ell^-_\kappa,\ell^+_1,\cdots,\ell^+_\kappa\}>\delta$. Note that some parameters have restricted spaces, such as $\lambda > 0$. In such cases, it is acceptable for the estimation to be close to the restrictions, and the percentage proximity criteria should not be calculated. We apply $\delta=0.01$ in most cases.
\item If the first-order optimality measure or the proximity criteria are not satisfied:
\begin{enumerate}[label*=\arabic*.]
\item If the first-order optimality measure is not close to zero, repeat the optimization process by returning to step 2 and choosing the next starting value. Continue this process until the first-order optimality criterion is met.
\item  If the first-order optimality is close to zero but the proximity criteria is not satisfied, proceed to step 3 and adjust the value of $r_1$ to $r_2$, where, $r_2>r_1$. If this second round of optimization still fails to satisfy the criteria, return to step 3 and use a much larger value of $r$, $r_3>r_2$, only for those parameters that do not meet the criteria.
\item Report the optimization as unsuccessful if all $s$ stored initial values fail to simultaneously meet both the first-order optimality and the proximity criteria.\footnote{In this paper we set $r_1$=1, $r_2$=2 and $r_3$=10, respectively.}
\end{enumerate}
\end{enumerate}
We use the algorithm described above to obtain maximum likelihood estimates for both the Aranda-Ordaz and Logit DDMS models. The optimization set-up is identical for both models, except in the logit case, where only the matrix of random numbers $C$ is used for the search of starting points. It is worth noting that we employ the same matrix $C$ for both models, ensuring that the starting points search is comparable across models.

\section{Monte Carlo Simulation}

In this section, we perform a Monte Carlo simulation study to compare the proposed Aranda-Ordaz approach to the traditional logit case following the contributions of \cite{maheu_identifying_2000, maheuvolatility2000}. More precisely, our analysis is divided into two model applications based on an empirically relevant set of parameters. In the first case, we analyze the in-sample state probabilities generated under duration uncertainty for both functions following a bull and bear market analysis as in \cite{maheu_identifying_2000}. For the second case, our study structure is quite similar; however, we explore the out-of-sample context, considering a conditional variance specification following \cite{maheuvolatility2000}. All codes were written in Matlab by the authors, and are available upon request.

\subsection{In-sample capabilities: the bull and bear market model}

{In the first Monte Carlo investigation, we focus on the in-sample predictive ability related to the transition probabilities of the proposed Aranda-Ordaz approach compared to the fixed link logit case. We consider the following model
\begin{equation}\label{incap}
Y_t=\mu_0(1-S_t)+\mu_1S_t+\big((1-S_t)\sigma_0+S_t\sigma_1\big)Z_t,\quad Z_t\sim N(0,1).
\end{equation}
In choosing the simulation scenario, we consider parameters reflecting the bull and bear market stylized facts. The bull (bear) market is characterized by positive (negative) mean returns and lower (higher) variance. For the transition probabilities parameters, we set $\gamma_2(i)>0$ for $i=0,1$, and this reflects that the probability of staying in the bull (bear) market increases as the duration increases. This dependence structure can be interpreted as a momentum effect, as discussed in \cite{maheu_identifying_2000, jp10, shibata}, among others.

We consider model \eqref{incap} with true duration $d_0=8$ and parameters $\mu_0=-0.5$, $\mu_1=1.5$, $\sigma_0=6$ and $\sigma_1=2$. The transition probabilities are given by \eqref{transprob} with parameters $\gamma_1(0)=-1.8$, $\gamma_2(0)=0.7$, $\gamma_1(1)=-0.8$ and $\gamma_2(1)=0.6$.\footnote{The parameters values scale refers to log returns multiplied by 100.} We estimate the model with duration $d\in\{4,6,8,10,12\}$ using both, the proposed Aranda-Ordaz and the fixed logit link. We generate time series of length 1,000 and discarded the first 200 observations as burn-in, yielding a final sample size of $n=800$. The experiment was replicated 1,000 times. For each time series we fit model \eqref{incap} using the logit and Aranda-Ordaz links. For each approach, we obtain the predictive, filtered and smoothed probabilities, denoted respectively by $P(S_t=i|\varphi_{t-1})$, $P({S_t} = i|\varphi _{t})$ and $P({S_t} = i|\varphi_{T})$ as in \cite{space}.

Tables \ref{probmape} and \ref{sup1} show the simulation results. Table \ref{probmape} presents the proportion of times the MAPE of the Aranda-Ordaz probability's MAPE is less than that of the logit's as well as the average MAPE difference for each probability type computed. Table \ref{sup1} shows the proportion of times the likelihood obtained using the proposed Aranda-Ordaz approach is greater than the likelihood obtained with the logit. From Table \ref{probmape}, we observe that when the model is misspecified (i.e. $\tau\neq\tau_0$), the Aranda-Ordaz approach yields more precise probabilities more often than the logit. The more distant is $\tau$ from $\tau_0$, the better the Aranda-Ordaz performs in comparison to the logit case.  Under the correct specification, however, the logit link outperforms the Aranda-Ordaz by a narrow margin, presenting the smallest of all MAPE differences, and a slightly superior proportion for smaller MAPE. On the other hand, the results presented in Table \ref{sup1} shows, on the other hand, that the Aranda-Ordaz produces a higher likelihood in the vast majority of cases, even in the correctly specified scenario.
\begin{table}[ht!]	
\caption{Proportion of times in which the MAPE obtained with the Aranda-Ordaz is smaller than the logit's and the difference between the respective average MAPE.}\label{probmape}
\centering
\renewcommand{\arraystretch}{1.3}
\vskip.3cm
\begin{tabular}{c|ccccc|ccccc}
	 \hline
\multirow{1}{*}{Probabilities}   &\multicolumn{5}{c|}{Proportion}&\multicolumn{5}{c}{Difference}\\
 \hline 
 \multicolumn{1}{c|}{$\tau$}& 4    & 6    & 8     & 10   & 12   &4&6&8&10&12 \\
       \hline
$P(S_t=i|\varphi_{t-1})$ & 69\% & 52\% &  43\% & 60\% & 78\% & 0.0245&  0.0057&  -0.0042& 0.0218 & 0.0334 \\
$P(S_t=i|\varphi_t)$     & 69\% & 53\% &  45\% & 60\% & 77\% & 0.0265&  0.0067&  -0.0030& 0.0453 & 0.0561 \\
$P(S_t=i|\varphi_T)$     & 79\% & 55\% &  41\% & 55\% & 72\% & 0.0622&  0.0138&  -0.0087& 0.1547 & 0.1700 \\
\hline
\end{tabular}
\end{table}
\begin{table}[h!]	
\centering
\caption{Frequency at which the likelihood values obtained with the  Aranda-Ordaz is superior to the one obtained with the logit.\vspace{.3cm}}\label{sup1}
\setlength{\tabcolsep}{12pt}
\renewcommand{\arraystretch}{1.5}
\fontsize{10pt}{9}\selectfont
\begin{tabular}{c|ccccc}
\hline
DGP & \multicolumn{5}{c}{Likelihood Value Superiority Frequency }\\
\hline
\multirow{2}{*}{$\tau_0=8$} & {94\%} & {96\%} & {91\%} & {88\%} & {91\%} \\
& ($\tau=4$)             & ($\tau=6$)             & ($\tau=8$)             & ($\tau=10$)             & ($\tau=12$)             \\
\hline
\end{tabular}
\end{table}

As an illustration, Figure \ref{fig10} presents the bear market filtered probabilities. For the 200th  path over 1{,}000 replications, and $\tau=4$ (misspecified case), both links depicted similar market phases with different probabilities in some periods, as depicted in Figure \ref{fig10_aa}. We also compare these estimates to the DGP in Figure \ref{fig10_bb}. From the plot, we observe that the estimated values follow closely the true probability values, with the accumulated advantage of the Aranda-Ordaz specification period by period. However, the overall results support the gain of the Aranda-Ordaz under misspecification.
\begin{figure}[ht!]	
	\centering
	\subfigure[Logit and Aranda-Ordaz]{\includegraphics[width=.45\textwidth]{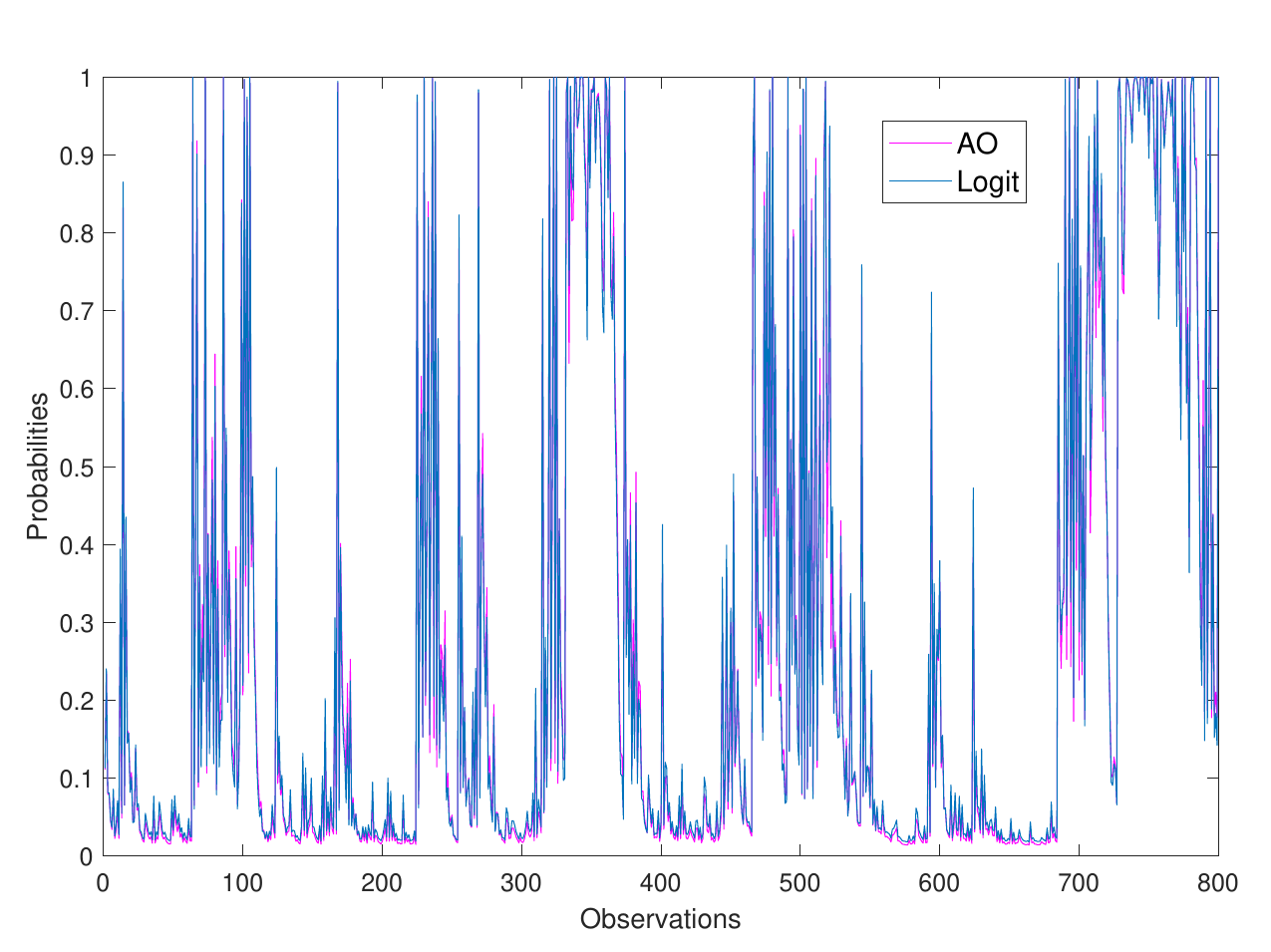}\label{fig10_aa}}
	\subfigure[DGP, Logit, and Aranda-Ordaz]{\includegraphics[width=.45\textwidth]{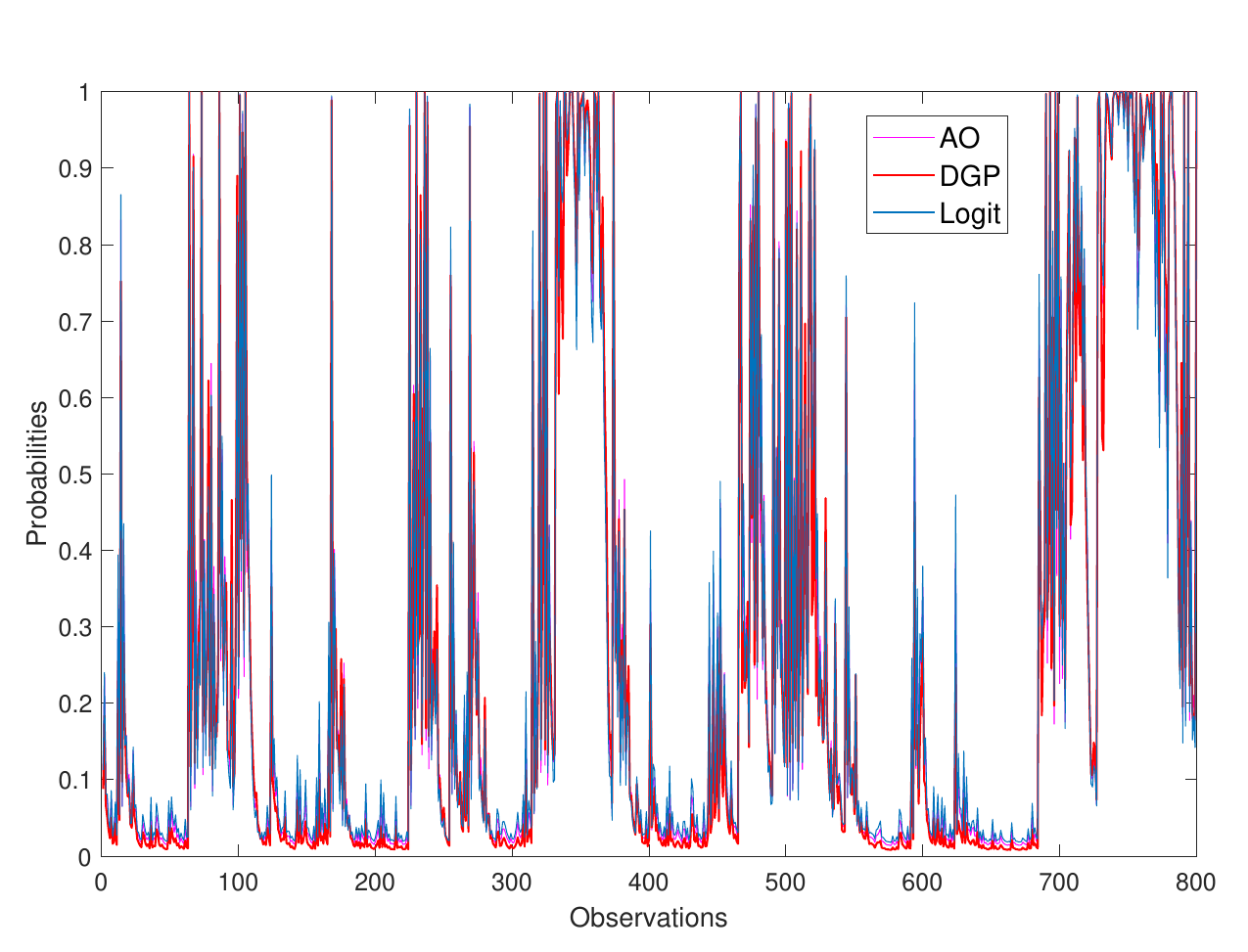}\label{fig10_bb}}
	\caption{Bear market filtered probabilities, for $\tau=4$.}\label{fig10}
\end{figure}
\subsection{Out-of-Sample capabilities: Volatility Forecasting}\label{forcap}
In this Section, we compare the proposed Aranda-Ordaz approach to the traditional logit approach in terms of out-of-sample capabilities. To provide grounds for comparison, we conduct a series of Monte Carlo simulations considering
\begin{align}\label{vol}
Y_t=\sigma\big(S_t,D(S_t)\big)Z_t,\quad\mbox{with}\quad
\sigma\big(S_t,D(S_t)\big)=\big(\omega(S_t)+\zeta(S_t)D(S_t)\big)^2,
\end{align}
where the latent state affects the level of volatility, $\omega(S_t)=\omega_0(1-S_t)+\omega_1S_t$, while the duration of the states, $D(S_t)$, affect the dynamics of volatility through $\zeta(S_t)D(S_t)$, where, $\zeta(S_t)=\zeta_0(1-S_t)+\zeta_1S_t$. $Z_t$ is assumed to be independent and identically distributed, following a standard normal distribution, and transition probabilities for the states $S_t$ are given by \eqref{transprob}.

We generated time series of size $n+10$, say $y_1,\cdots,y_{n+10}$ for values of $\tau_0\in\{15,25,35\}$ and parameters $\omega_0=1.0$, $\omega_1=1.3$, $\zeta_0=-0.01$ and $\zeta_1=0.02$ with transition probabilities given by $\gamma_1(0)=1.0$, $\gamma_2(0)=0.1$, $\gamma_1(1)=1.3$ and $\gamma_2(1)=-0.01$, which are similar to those used in \cite{maheuvolatility2000}. The last 10 values were reserved for out-of-sample forecasting purposes with $n$=1000 after discarting the first 200 observations as burn-in. For each scenario, we estimate the model using $y_1,\cdots,y_n$ with the logit and the Aranda-Ordaz link functions for $\tau\in\{\tau_0-10,\tau_0-5, \tau_0, \tau_0+5,\tau_0+10\}$.  Next, for each estimated model we obtain $h$-steps ahead forecast for forecasting horizons $h\in\{1,\cdots,10\}$. Let $\hat \sigma^2_{n+1}, \cdots, \hat \sigma^2_{n+10}$ and $\tilde \sigma^2_{n+1}, \cdots, \tilde \sigma^2_{n+10}$ denote the forecasted values of $\sigma^2_{n+1},\cdots,\sigma^2_{n+10}$ using the Aranda-Ordaz and the logit, respectively. For each horizon $h\in\{1,\cdots,10\}$ and each method, we calculate the forecasting mean absolute percentage error (MAPE), given by
\begin{equation*}
\mathrm{MAPE}_{\mathrm{AO}}(h)=\frac1h\sum_{k=1}^h \bigg|\frac{\hat \sigma^2_{n+k}-\sigma^2_{n+k}}{\sigma^2_{n+k}}\bigg|,\qquad\mbox{and}\qquad
\mathrm{MAPE}_{\mathrm{logit}}(h)=\frac1h\sum_{k=1}^h \bigg|\frac{\tilde \sigma^2_{n+k}-\sigma^2_{n+k}}{\sigma^2_{n+k}}\bigg|.
\end{equation*}
We replicate the experiment 1,000 times. To simplify the exposition, in Table \ref{difmape} we present the difference between the average MAPE of the proposed Aranda-Ordaz approach and the logit, for each forecast horizon, that is, $D(h):=\mathrm{MAPE}_{\mathrm{logit}}(h)- \mathrm{MAPE}_{\mathrm{AO}}(h)$, for $h\in\{1,\cdots,10\}$.
\begin{table}[h!]	
\caption{ Predictive ability of the logit and the Aranda-Ordaz links in a simulated variance forecasting exercise. Presented are the difference between the MAPE  of  the proposed Aranda-Ordaz approach and the logit. Positive values favor the Aranda-Ordaz specification while negative values stand for the logit case. \vspace{.5cm}}\label{difmape}
\centering
\renewcommand{\arraystretch}{1.4}
\begin{adjustbox}{max width=\textwidth}
\begin{tabular}{c|c|cccccccccc}
\hline
 DGP        & Model    & \multicolumn{10}{c}{Forecasting horizon ($h$)}  \\ \cline{3-12}
$\tau_0$&          $\tau$                  & {1}      & {2}      & {3}      & {4}      & {5}      & {6}      & {7}      & {8}      & {9}      & {10} \\
\hline
\multirow{5}{*}{$15$} & {$5$}  & {0.0232} & {0.0173} & {0.0151} & {0.0153} & {0.0146} & {0.0136} & {0.0128} & {0.0122} & {0.0115} & {0.0104} \\
& {$10$} & {0.0749} & {0.0614} & {0.0533} & {0.0463} & {0.0407} & {0.0357} & {0.0323} & {0.0293} & {0.0264} & {0.0235} \\
& {$15$} & {0.1295} & {0.1001} & {0.0840} & {0.0720} & {0.0637} & {0.0574} & {0.0516} & {0.0467} & {0.0430} & {0.0399} \\
& {$20$} & {0.0328} & {0.0278} & {0.0236} & {0.0197} & {0.0163} & {0.0141} & {0.0121} & {0.0102} & {0.0093} & {0.0091} \\
& {$25$} & {0.0482} & {0.0399} & {0.0319} & {0.0262} & {0.0217} & {0.0182} & {0.0155} & {0.0133} & {0.0114} & {0.0096} \\
\hline
\multirow{5}{*}{$25$} & {$15$} & {0.4406} & {0.3994} & {0.3713} & {0.3505} & {0.3312} & {0.3129} & {0.2972} & {0.2833} & {0.2712} & {0.2608} \\
& {$20$} & {0.0959} & {0.0909} & {0.0860} & {0.0829} & {0.0807} & {0.0788} & {0.0754} & {0.0724} & {0.0701} & {0.0683} \\
& {$25$} & {0.1750} & {0.1578} & {0.1477} & {0.1399} & {0.1350} & {0.1293} & {0.1236} & {0.1188} & {0.1144} & {0.1106} \\
& {$30$} & {0.4244} & {0.3862} & {0.3602} & {0.3375} & {0.3180} & {0.3020} & {0.2874} & {0.2728} & {0.2600} & {0.2481} \\
& {$35$} & {0.2614} & {0.2354} & {0.2168} & {0.2056} & {0.1960} & {0.1876} & {0.1798} & {0.1726} & {0.1661} & {0.1594} \\
\hline
\multirow{5}{*}{$35$} & {$25$} & {0.6613} & { 0.6415} & {0.6260} & {0.6135} & {0.6018 } & {0.5902 } & {0.5806} & {0.5723} & {0.5666} & {0.5623}\\
& {$30$} & {1.3123} & {1.2725} & {1.2461} & {1.2205} & {1.1912} & {1.1630} & {1.1386} & {1.1173} & {1.0966} & {1.0771} \\
& {$35$} & {1.1140} & {1.0849} & {1.0577} & {1.0303} & {1.0026} & {0.9816} & {0.9653} & {0.9517} & {0.9371} & {0.9227} \\
& {$40$} & {1.1296} & {1.0663} & {1.0188} & {0.9727} & {0.9341} & {0.9062} & {0.8848} & {0.8684} & {0.8528} & {0.8353} \\
& {$45$} & {1.0551} & {1.0083} & {0.9653} & {0.9238} & {0.8887} & {0.8585} & {0.8373} & {0.8161} & {0.7945} & {0.7707} \\
 \hline
\end{tabular}
\end{adjustbox}
\end{table}
\FloatBarrier
Table \ref{difmape} only contains positive values indicating that, on average, forecasting using the Aranda-Ordaz is advantageous regardless the true value of $\tau_0$ and the duration $\tau$ used in the estimation procedure. It is interesting to notice that the smallest values of the difference $D(h)$ are usually not obtained when $\tau=\tau_0$, which may be a consequence of model complexity, indicating that even when the correct duration is used, the model based on the logit is not capable to provide the ``best possible'' forecasts. Regarding the magnitude of $\tau_0$, the higher the duration, the bigger the difference $D(h)$ for all forecasting horizons. The smallest differences $D(h)$ are obtained for $\tau_0=15$, $\tau=5$, ranging between about 10\% and 23\%, while the overall higher are obtained for $\tau_0=35$ and $\tau=30$, ranging from over 100\% to a bit over 130\%.
%

%
The results on Table \ref{difmape} show that, on average, applying the Aranda-Ordaz link is advantageous over the logit even when the duration and the link function are correctly specified. One question that remains is how often (if at all) the likelihood of the proposed Aranda-Ordaz approach is higher than the logit. To shed some light into this issue, under the same DGP as in Section \ref{forcap}, we compare the likelihood obtained using the proposed Aranda-Ordaz approach and the logit.

Table \ref{sup} presents the frequency at which the log-likelihood obtained using the Aranda-Ordaz link is superior to the one obtained with the logit. The first column presents the true value $\tau_0$ applied in the data generating process, while the value of $\tau$ used in the estimation procedure is displayed in parenthesis. We observe that in all scenarios considered in the simulation, using the Aranda-Ordaz link yields superior likelihood in average, including the case where the link and duration are correctly specified. This result is expected since the Aranda-Ordaz link is more flexible than the logit. The frequency at which the likelihood is greater for the Aranda-Ordaz ranges from 62\% to 91\%. Interestingly, the higher the true value of $\tau$, the greater this frequency is.
\begin{table}[h!]	
	\centering
	\caption{  Frequency at which the Aranda-Ordaz link likelihood value is superior to the one obtained with the logit link in 1{,}000 simulated trials.\vspace{.3cm}}\label{sup}
\setlength{\tabcolsep}{10pt}
\renewcommand{\arraystretch}{1.5}
\begin{adjustbox}{max width=\textwidth}
\fontsize{10pt}{9}\selectfont
\begin{tabular}{c|ccccc}
\hline
$\tau_0$ (DGP) & \multicolumn{5}{c}{Log-Likelihood Value Superiority Frequency }\\
\hline
\multirow{2}{*}{$15$} & {62\%} & {70\%} & {63\%} & {62\%} & {68\%} \\
& ($\tau=5$)             & ($\tau=10$)             & ($\tau=15$)             & ($\tau=20$)             & ($\tau=25$)             \\ \hline
\multirow{2}{*}{$25$} & {89\%} & {76\%} & {78\%} & {87\%} & {68\%} \\
& ($\tau=15$)             & ($\tau=20$)             & ($\tau=25$)             & ($\tau=30$)             & ($\tau=35$)             \\ \hline
\multirow{2}{*}{$35$} & {87\%} & {91\%} & {88\%} & {82\%} & {84\%} \\
& ($\tau=25$)             & ($\tau=30$)             & ($\tau=35$)             & ($\tau=40$)             & ($\tau=45$)             \\ \hline
\end{tabular}
\end{adjustbox}
\end{table}

\section{Empirical Exercise}

In this section, we present a real data application of the proposed methodology considering the log returns of daily closing S\&P500 index from January 2nd 2015 to January 2nd 2020, yielding a sample size $n=1{,}259$, as seen in Figure \ref{fig2}. The descriptive statistics is presented in Table \ref{stats}. We observe a mean very close to zero and a small standard deviation, with an annualized value of 0.1344 (0.0085 $\times \sqrt {250}$ ). The Jarque-Bera (JB) normality test rejects the null hypothesis of an underlying normal distribution, which is further corroborated by the kurtosis, considerably higher than the normal distribution's. The Lagrange Multiplier and Ljung-Box tests are highly significant, suggesting ARCH effects in the log returns.
\begin{figure}[h]	
	\centering
	\subfigure[S\&P500 Index]{\includegraphics[width=.48\textwidth]{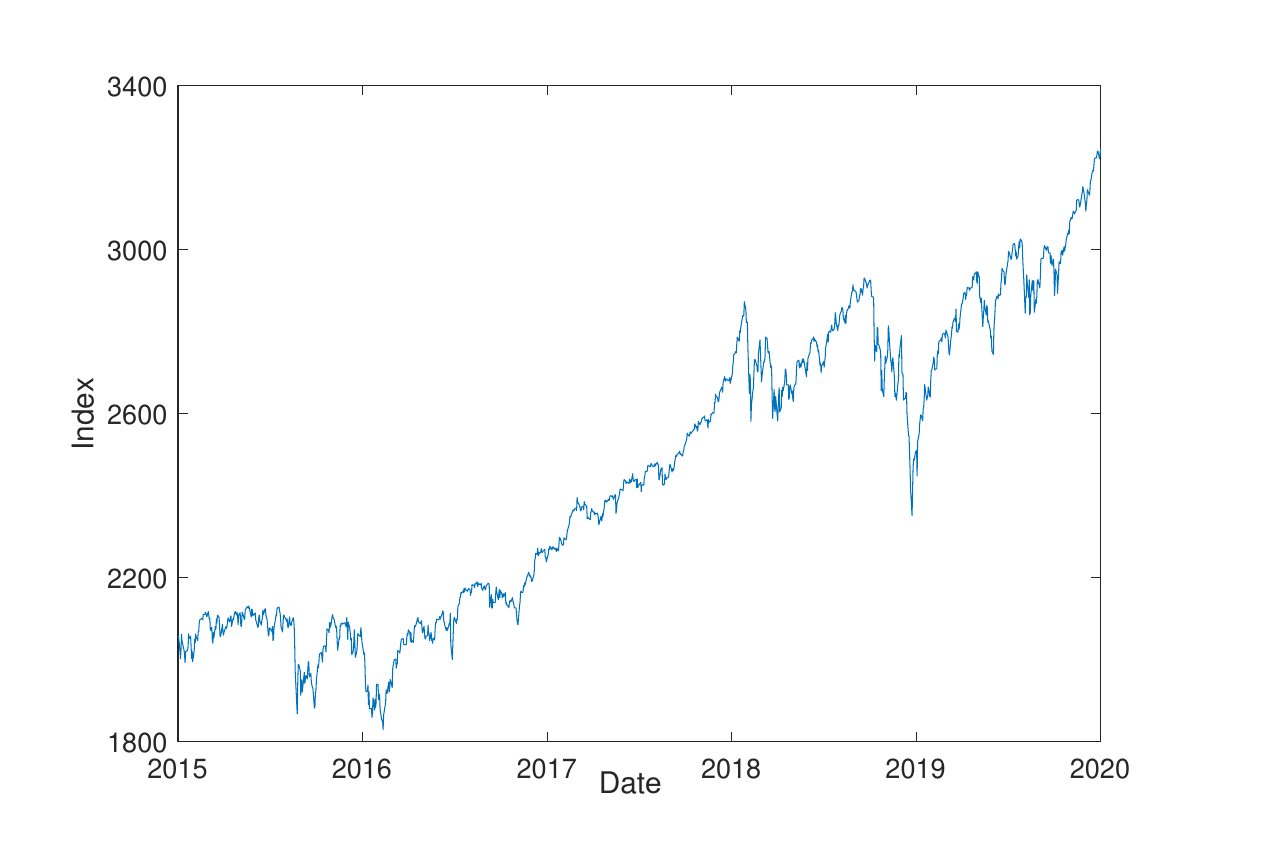}}
	\subfigure[S\&P500 Log Returns]{\includegraphics[width=.43\textwidth]{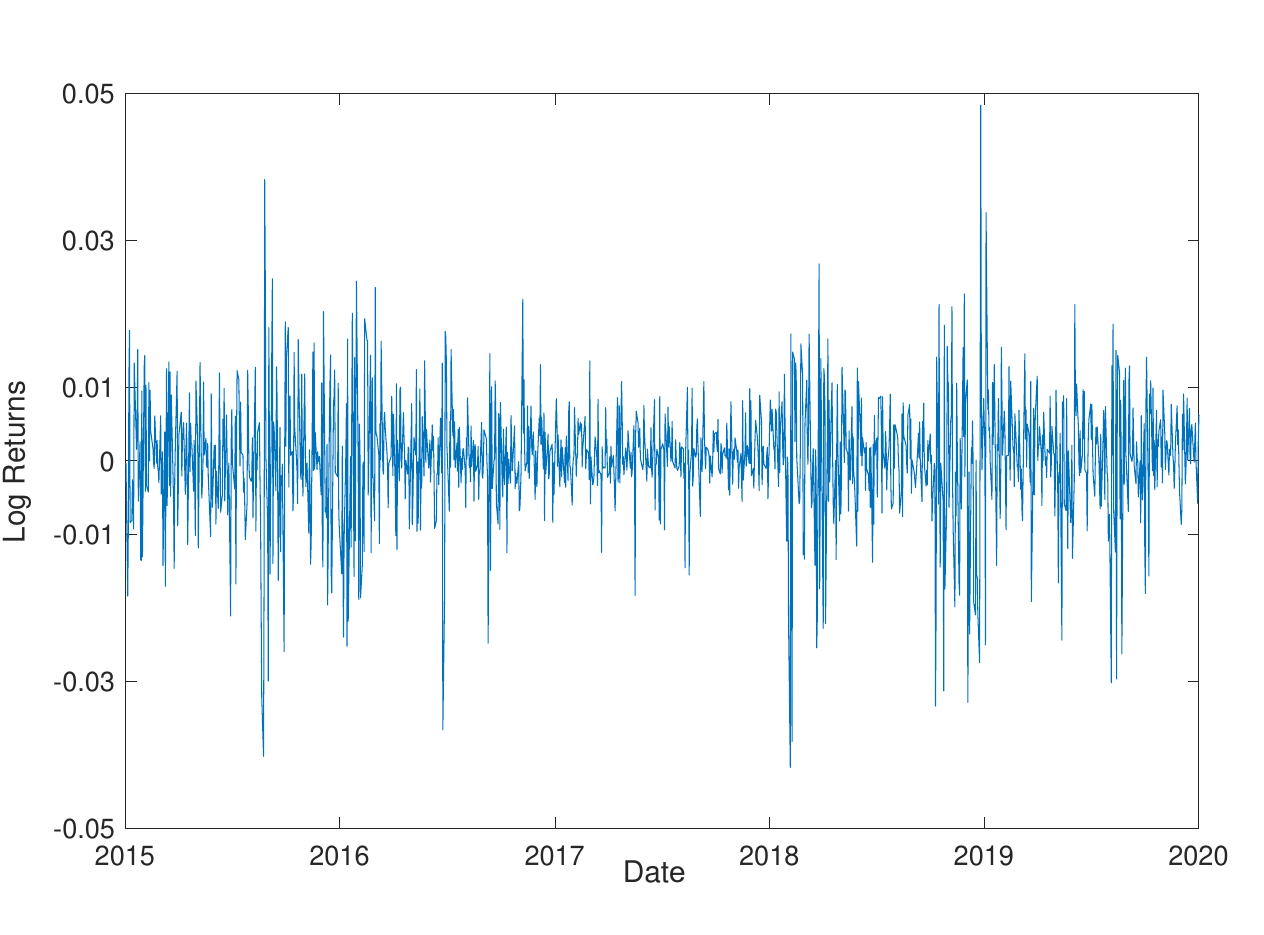}}
	\caption{Time series plot of the S\&P500 (a) index and (b) log-returns.}	 \label{fig2}
\end{figure}
\begin{table}[h!]	
	\centering
	\caption{Summary statistics for the S\&P500 daily log-returns. JB refers to the Jarque-Bera normality test, LM stands for the Lagrange Multiplier test for ARCH effects in the demeaned returns, while $Q^2_k$ is the corresponding Ljung-Box statistic on the squared demeaned returns considering $k$ lags, respectively.  \vspace{.3cm}}
	\label{stats}
\begin{adjustbox}{max width=\textwidth}
\begin{tabular}{c|c|c|c|c|c|c|c|c| c}
\hline
$n$ & {Mean} &{Std. Dev.} & {Max.} & {Min.} & {Skewness} & {Kurtosis} &{JB} &{LM(8)} & {$Q^2_8$} \\
\hline
{1259} & {0.0004} & {0.0085} & {0.0484} & {-0.0418} & {-0.5260} & {6.8360} & {$829.9^{*}$} & {$1899.8^{*}$} & {$373.8^{*}$} \\
\hline
\multicolumn{7}{l}{\footnotesize ${}^\ast$ indicates the rejection of the null hypotheses at 1\% significance level.} &\multicolumn{2}{c}{ }\\
\end{tabular}
\end{adjustbox}
\end{table}	
Our goal is to conduct an out-of-sample forecasting exercise considering a 1-day step ahead volatility forecast through an expanding window using 443 daily observations, starting on April 1st, 2018. During this period, the S\&P500 presented evidence of intra-state fluctuations, such as bear rallies (positive sub-trend) and bull corrections (negative sub-trend), encompassing a primary bull market state as described by the empirical literature \citep[see, for example,][for stock market cycles] {MAHEU2021102091}.  We use this period of uncertainty and volatility to run a point forecasting exercise to assess the performance of a single Aranda-Ordaz DDMS model compared to benchmark models.
\subsection{Realized measures}
A well-known problem in financial econometrics is that volatility is a latent variable and, therefore, cannot be directly observed, turning the evaluation of the predictive power of different approaches problematic. In this section, we consider the 5-minute intra-day quotes as an alternative source to proxy volatility. More specifically, besides the traditional realized variance \citep[RV -- see][]{h1}, we also consider alternative realized measures robust to jump and microstructure noise to proxy the true volatility, namely the bipower variation \citep[BV -- see][]{barndorff-nielsen_power_2004}, MinRV and  MedRV  \citep[][]{andersen_jump-robust_2012}.

In general, these measures are regarded as better alternatives to proxy the true volatility compared to the squared daily return \citep[see, for instance,][among others.]{h2, h3}. Their definition is as follows:
\begin{align*}
\mathrm{RV}_t&:=\sum_{i = 1}^m r_{i,t}^2,\\
\mathrm{BV}_t&:=\frac{\pi }{2}\left[ \frac{m}{m - 1} \right]\sum_{i = 1}^{m - 1} |r_{i,t}r_{i + 1,t}|,\\
\mathrm{MinRV}_t&:=\frac{\pi}{\pi  - 2}\left[ \frac{m}{m - 1} \right]\sum_{i = 1}^{m - 1}\min\bigl\{|r_{i,t}|,|r_{i + 1,t}|\big\}^2,\\
\mathrm{MedRV}_t&:=\frac{\pi }{{6 - 4\sqrt 3  + \pi }}\left[ \frac{m}{m - 2}\right]\sum_{i = 2}^{m - 1} \mathrm{med}\big\{|r_{i - 1,t}|,|r_{i,t}|,|r_{i + 1,t}|\big\}^2,
\end{align*}
where, $r_{i,t}$ is the $i$th high-frequency return of day $t$ with $m=78$ cases. The 5 minutes intra-day quotes range from 9.30 AM to 4.00 PM, and the time series were obtained from First Rate Data website {\color{blue}\href{https://firstratedata.com}{https://firstratedata.com}}.
\subsection{Robust loss function}
Typically, volatility proxies are used to evaluate volatility forecasts. However, these proxies are estimates of the integrated variance and, as such, they are imperfect. \cite{patton_volatility_2011} defines a sense of robustness for loss functions in ranking volatility forecasts, and based on this concept, derived a general class of loss functions that are robust in that sense. Letting $\bar\sigma^2$ the volatility proxy and $\hat\sigma^2$ the volatility forecasts, we consider three loss functions to evaluate volatility forecasts that are members of \cite{patton_volatility_2011}'s class, namely, the MSE, the QLIKE,  and a measure denote hereafter by RLF, given by
\begin{align*}
\mathrm{MSE}(\bar\sigma^2,\hat\sigma^2)&:=\frac{(\bar\sigma^2-\hat\sigma^2)^2}2, \quad \mathrm{QLIKE}(\bar\sigma^2,\hat\sigma^2):=\frac{\bar\sigma^2}{\hat\sigma^2}-\log\biggl(\frac{\bar\sigma^2}{\hat\sigma^2}\bigg)-1,\\
\mathrm{RLF}(\bar\sigma^2,\hat\sigma^2)&:=\hat\sigma^2 - \bar\sigma^2\bigg[\log\biggl( \frac{\bar\sigma^2 }{\hat\sigma^2}\bigg)-1\bigg].
\end{align*}
Observe that \cite{patton_volatility_2011}'s MSE and QLIKE losses differ from the usually applied loss functions of the same name.
\subsection{Results}
We are considering the volatility model described in Equation \eqref{vol}, which was previously examined in a Monte Carlo simulation study. To ensure a reliable comparison between the Logit and Aranda-Ordaz links, we are using the same parameter setup to optimize the likelihood functions of the model. This study is constructed using a Model Confidence Set and a pairwise test including Garch-type models .
\subsubsection*{Parameter Estimates}
Before comparing the predictive performance of the Aranda-Ordaz model with the benchmark specifications, we first examine the $\lambda$ estimates at each step during the expanding window estimation process for the DDMS A-O model with $\tau=5$. In Figure \ref{fig4_a}, we present the estimated values of $\lambda$ as a function of the expanding windows time stamps. The lowest estimated value is 0.0109 and the highest is 5.8403. The pattern observed for the estimated values of $\lambda$ shed light on the complexity of the model specification when applied to the data at hand. This complex dynamic cannot be accounted by the model with fixed link.



In this study, we deliberately chose a shorter duration value for the Aranda-Ordaz specification, setting it apart from the usual logit alternatives used as benchmarks. Our aim is to adopt a model that not only addresses the uncertainties associated with duration but also balances model complexity. As previously emphasized, estimating these models becomes increasingly challenging as longer durations are considered. Our objective is to determine whether a short duration model with a flexible link can outperform traditional logit-based models even considering different values for the duration.
\begin{figure}[h!]	
	\centering
{\includegraphics[width=.38\textwidth]{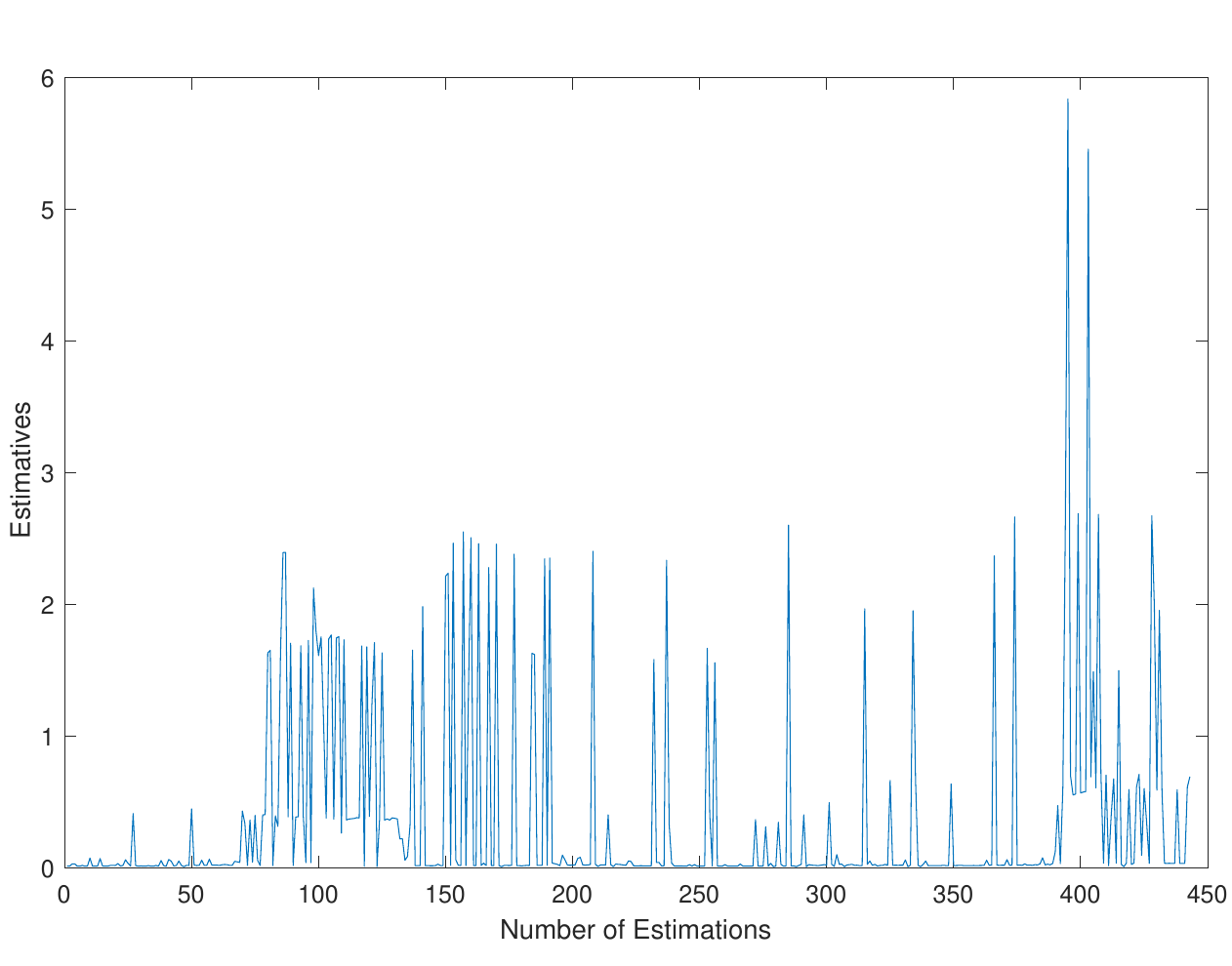}\label{fig4_a}}
	\caption{Estimated values of $\lambda$ for the DDMS A-O model with $\tau=5$. }	 \label{fig4}
\end{figure}
\subsubsection*{Forecasting Results -- Model Confidence Set (MCS)}
Table \ref{tunk3} provides the average MSE, RLF, and the QLIKE obtained from the forecasted values of $\sigma_{n+1}^2$, along with their corresponding realized measures. In gray are the models with superior out-of-sample performance, as determined by the Model Confidence Set (MCS) approach outlined by \cite{hansen2011}, with an 80$\%$ significance level. The best models in their respective categories are indicated in bold. In general, the DDMS A-O model with $\tau = 5$ is the best-performing model in the MCS methodology for every combination of loss function and realized measures. We set the benchmark as the logit models for $\tau\in\{ 5, 15, 25\}$, and also combining these models through a simple mean approach, since duration is unknown in this empirical exercise. We also tested other durations for the Aranda-Ordaz specification against the benchmark model. For the Aranda-Ordaz model with $\tau=15$, we observed partially similar results, but the model estimation tends to become more complex for longer durations. Model complexity also affects the logit specification. In the case $\tau=25$, convergence in the optimization procedure failed on the dates 06/29/2018 and 11/09/2018, resulting in a final out-of-sample size study of 441 observations.
\begin{table}[h!]	
\caption{ Average MSE, RLF and QLIKE values of the forecasting values of $\sigma _{n + 1}^2$ and realized measures. Gray cells indicate the set of models with the best out-of-sample performance obtained using the MCS approach \citep{hansen2011} at the 80\% significance level. In bold, the best model (first position in the MCS rank) in each case.
	\vspace{.5cm}}\label{tunk3}
\centering
\renewcommand{\arraystretch}{1.3}
\begin{adjustbox}{max width=\textwidth}
\begin{tabular}{c|llll|llll|llll}
	\hline
& \multicolumn{4}{c|}{} & \multicolumn{4}{c|}{} & \multicolumn{4}{c}{} \\
& \multicolumn{4}{c|}{\multirow{-2}{*}{MSE  ($\times10^8$)  }}  & \multicolumn{4}{c|}{\multirow{-2}{*}{RLF  ($\times10^4$) }} & \multicolumn{4}{c}{\multirow{-2}{*}{QLIKE }} \\ \cline{2-13}
& \multicolumn{1}{c}{}  & \multicolumn{1}{c}{}
   & \multicolumn{1}{c}{} & \multicolumn{1}{c|}{}    & \multicolumn{1}{c}{}  & \multicolumn{1}{c}{} & \multicolumn{1}{c}{}   & \multicolumn{1}{c|}{}    & \multicolumn{1}{c}{} & \multicolumn{1}{c}{}
   & \multicolumn{1}{c}{}    & \multicolumn{1}{c}{}   \\	\multirow{-4}{*}{Models}     & \multicolumn{1}{c}{\multirow{-2}{*}{MedRV}}
& \multicolumn{1}{c}{\multirow{-2}{*}{MinRV}} & \multicolumn{1}{c}{\multirow{-2}{*}{BV}}
& \multicolumn{1}{c|}{\multirow{-2}{*}{RV}}                  & \multicolumn{1}{c}{\multirow{-2}{*}{MedRV}}                & \multicolumn{1}{c}{\multirow{-2}{*}{MinRV}}                & \multicolumn{1}{c}{\multirow{-2}{*}{BV}}                   & \multicolumn{1}{c|}{\multirow{-2}{*}{RV}}                  & \multicolumn{1}{c}{\multirow{-2}{*}{MedRV}}                & \multicolumn{1}{c}{\multirow{-2}{*}{MinRV}}                                       & \multicolumn{1}{c}{\multirow{-2}{*}{BV}}   & \multicolumn{1}{c}{\multirow{-2}{*}{RV}}                  \\ \hline
& \cellcolor[HTML]{EFEFEF}   & \cellcolor[HTML]{EFEFEF}             & \cellcolor[HTML]{EFEFEF} & \cellcolor[HTML]{EFEFEF}      & \cellcolor[HTML]{EFEFEF}       & \cellcolor[HTML]{EFEFEF} & \cellcolor[HTML]{EFEFEF}       & \cellcolor[HTML]{EFEFEF}                                   & \cellcolor[HTML]{EFEFEF}   & \cellcolor[HTML]{EFEFEF}{\color[HTML]{000000} }     & \cellcolor[HTML]{EFEFEF}    & \cellcolor[HTML]{EFEFEF}    \\	 \multirow{-2}{*}{A-O $\tau=5$}        & \multirow{-2}{*}{\cellcolor[HTML]{EFEFEF}\textbf{0.52207}} & \multirow{-2}{*}{\cellcolor[HTML]{EFEFEF}\textbf{0.52328}} & \multirow{-2}{*}{\cellcolor[HTML]{EFEFEF}\textbf{0.51303}} & \multirow{-2}{*}{\cellcolor[HTML]{EFEFEF}\textbf{0.51326}} & \multirow{-2}{*}{\cellcolor[HTML]{EFEFEF}\textbf{0.26888}} & \multirow{-2}{*}{\cellcolor[HTML]{EFEFEF}\textbf{0.27465}} & \multirow{-2}{*}{\cellcolor[HTML]{EFEFEF}\textbf{0.26271}} & \multirow{-2}{*}{\cellcolor[HTML]{EFEFEF}\textbf{0.25132}} & \multirow{-2}{*}{\cellcolor[HTML]{EFEFEF}\textbf{0.49920}} & \multirow{-2}{*}{\cellcolor[HTML]{EFEFEF}{\color[HTML]{000000} \textbf{0.51841}}} & \multirow{-2}{*}{\cellcolor[HTML]{EFEFEF}\textbf{0.48078}} & \multirow{-2}{*}{\cellcolor[HTML]{EFEFEF}\textbf{0.44292}} \\
&          & \cellcolor[HTML]{FFFFFF}      & \cellcolor[HTML]{FFFFFF} & \cellcolor[HTML]{FFFFFF}      & & \cellcolor[HTML]{FFFFFF} & \cellcolor[HTML]{FFFFFF}    & \cellcolor[HTML]{FFFFFF}    & & {\color[HTML]{000000} }    & \cellcolor[HTML]{FFFFFF}   & \cellcolor[HTML]{FFFFFF}     \\	 \multirow{-2}{*}{Logit  $\tau=5$}     & \multirow{-2}{*}{3.29668}        & \multirow{-2}{*}{\cellcolor[HTML]{FFFFFF}3.29282}          & \multirow{-2}{*}{\cellcolor[HTML]{FFFFFF}3.26904}     & \multirow{-2}{*}{\cellcolor[HTML]{FFFFFF}3.28132}          & \multirow{-2}{*}{0.49674} & \multirow{-2}{*}{\cellcolor[HTML]{FFFFFF}0.50567}      & \multirow{-2}{*}{\cellcolor[HTML]{FFFFFF}0.49038}          & \multirow{-2}{*}{\cellcolor[HTML]{FFFFFF}0.47641}     & \multirow{-2}{*}{0.61005}     & \multirow{-2}{*}{{\color[HTML]{000000} 0.63314}} & \multirow{-2}{*}{\cellcolor[HTML]{FFFFFF}0.59097}     & \multirow{-2}{*}{\cellcolor[HTML]{FFFFFF}0.54907}          \\
& \cellcolor[HTML]{EFEFEF}    & \cellcolor[HTML]{EFEFEF}  & \cellcolor[HTML]{EFEFEF} & \cellcolor[HTML]{EFEFEF} &   & \cellcolor[HTML]{FFFFFF}   & \cellcolor[HTML]{FFFFFF}      & \cellcolor[HTML]{FFFFFF}&  & {\color[HTML]{000000} } & \cellcolor[HTML]{FFFFFF}    & \cellcolor[HTML]{FFFFFF}   \\
\multirow{-2}{*}{Logit  $\tau=15$}    & \multirow{-2}{*}{\cellcolor[HTML]{EFEFEF}2.25275}          & \multirow{-2}{*}{\cellcolor[HTML]{EFEFEF}2.25101}  & \multirow{-2}{*}{\cellcolor[HTML]{EFEFEF}2.23970}    & \multirow{-2}{*}{\cellcolor[HTML]{EFEFEF}2.25431}   & \multirow{-2}{*}{0.39282}                                  & \multirow{-2}{*}{\cellcolor[HTML]{FFFFFF}0.39771}  & \multirow{-2}{*}{\cellcolor[HTML]{FFFFFF}0.38626}          & \multirow{-2}{*}{\cellcolor[HTML]{FFFFFF}0.37644}   & \multirow{-2}{*}{0.61748}                                  & \multirow{-2}{*}{{\color[HTML]{000000} 0.63612}}     & \multirow{-2}{*}{\cellcolor[HTML]{FFFFFF}0.59773}          & \multirow{-2}{*}{\cellcolor[HTML]{FFFFFF}0.55795}          \\	& \cellcolor[HTML]{EFEFEF}    & \cellcolor[HTML]{EFEFEF}    & \cellcolor[HTML]{EFEFEF} & \cellcolor[HTML]{EFEFEF} & & \cellcolor[HTML]{FFFFFF} & \cellcolor[HTML]{FFFFFF} & \cellcolor[HTML]{FFFFFF}   &  & {\color[HTML]{000000} } & \cellcolor[HTML]{FFFFFF}  & \cellcolor[HTML]{FFFFFF}     \\
\multirow{-2}{*}{Logit  $\tau=25$}    & \multirow{-2}{*}{\cellcolor[HTML]{EFEFEF}1.81357}          & \multirow{-2}{*}{\cellcolor[HTML]{EFEFEF}1.83117}          & \multirow{-2}{*}{\cellcolor[HTML]{EFEFEF}1.81060}          & \multirow{-2}{*}{\cellcolor[HTML]{EFEFEF}1.80456}          & \multirow{-2}{*}{0.47412}                                  & \multirow{-2}{*}{\cellcolor[HTML]{FFFFFF}0.48266}          & \multirow{-2}{*}{\cellcolor[HTML]{FFFFFF}0.46891}          & \multirow{-2}{*}{\cellcolor[HTML]{FFFFFF}0.45562}          & \multirow{-2}{*}{0.71623}                                  & \multirow{-2}{*}{{\color[HTML]{000000} 0.73603}}    & \multirow{-2}{*}{\cellcolor[HTML]{FFFFFF}0.69561}          & \multirow{-2}{*}{\cellcolor[HTML]{FFFFFF}0.65187}          \\
& \cellcolor[HTML]{EFEFEF}    & \cellcolor[HTML]{EFEFEF}   & \cellcolor[HTML]{EFEFEF} & \cellcolor[HTML]{EFEFEF}  & & \cellcolor[HTML]{FFFFFF}    & \cellcolor[HTML]{FFFFFF}   & \cellcolor[HTML]{FFFFFF} &     & {\color[HTML]{000000} }  & \cellcolor[HTML]{FFFFFF}     & \cellcolor[HTML]{FFFFFF}  \\
\multirow{-2}{*}{Logit Combination} & \multirow{-2}{*}{\cellcolor[HTML]{EFEFEF}1.38849}          & \multirow{-2}{*}{\cellcolor[HTML]{EFEFEF}1.39249}     & \multirow{-2}{*}{\cellcolor[HTML]{EFEFEF}1.37394}          & \multirow{-2}{*}{\cellcolor[HTML]{EFEFEF}1.38088}    & \multirow{-2}{*}{0.40924}     & \multirow{-2}{*}{\cellcolor[HTML]{FFFFFF}0.41643}   & \multirow{-2}{*}{\cellcolor[HTML]{FFFFFF}0.40182}          & \multirow{-2}{*}{\cellcolor[HTML]{FFFFFF}0.38822}    & \multirow{-2}{*}{0.65800}                                  & \multirow{-2}{*}{{\color[HTML]{000000} 0.67846}}    & \multirow{-2}{*}{\cellcolor[HTML]{FFFFFF}0.63570}          & \multirow{-2}{*}{\cellcolor[HTML]{FFFFFF}0.58944}   \\ \hline
\end{tabular}
\end{adjustbox}
\end{table}
\FloatBarrier
\subsubsection*{Comparison to Garch-type Models}
Our analysis also include the traditional Garch-type models in a pairwise evaluation. Based on \cite{haas_new_2004}, we use a plain vanilla specification, which is given by
\begin{align*}
Y_t&=\sigma_{k,t}Z_t,\\
\sigma_{k,t}^2 &= \omega_k + \alpha_kY_{t-1}^2+\beta_k\sigma_{k,t-1}^2,
\end{align*}
where $k$ is the number of regimes. We considered $k=1$, the traditional model, and $k=2$, the two-regime model. In both models $Z_t$'s are i.i.d. $N(0,1)$. Despite many variations of this modeling approach, our objective is to compare the performance of the links relative to a simple and useful model.
\subsubsection*{Fixed Predictive Accuracy}
Table \ref{tunk4} presents the $t$-statistics from Diebold-Mariano-West tests of equal predictive accuracy between the benchmark Garch model and the DDMS models for $\tau\in\{5,15,25\}$, considering the same three losses and four realized variances as in the MCS example. A positive $t$-statistic indicates that the DDMS models' forecasts produced a larger average loss than the Garch models. A $t$-statistic greater than 1.65 and 1.96 in absolute value indicates a rejection of the null hypothesis of equal predictive accuracy at the 0.10 and 0.05 levels. We also report $p$-values in parentheses for easy reference.

In general, we observe that the null hypotheses of equal predictability for the Aranda-Ordaz DDMS and Garch models cannot be rejected at the 0.10 level in most of cases for different combinations of loss and realized variances. For the logit specification, the null rejection is uniformly observed is most of cases a 0.01 level. The only exception is found for the logit $\tau=15$ for the MSE. However, these results present a $p$-value lower than the Aranda-Ordaz case for the same loss and realized variance configuration.
\begin{table}[h!]	
\caption{Results from the Diebold-Mariano-West test applied to the Garch model. Presented are the $t$-statistics from Diebold-Mariano-West test along with their $p$-value in parentheses.
\vspace{.5cm}}\label{tunk4}
\centering
\renewcommand{\arraystretch}{1.3}
\begin{adjustbox}{max width=\textwidth}
\begin{tabular}{c| c|c|rc|rc|rc|rc}
\hline
\multicolumn{2}{c|}{Models}& Loss  & \multicolumn{2}{c|}{MedRV}  & \multicolumn{2}{c|}{MinRV}  & \multicolumn{2}{c|}{BV}  & \multicolumn{2}{c}{RV}     \\
\hline
\multirow{15}{*}{Garch} & \multirow{3}{*}{\begin{tabular}{c} A-O\\  $\tau=5$ \end{tabular}} & MSE & -0.01 &  (0.992) & -0.32  & (0.749)  & -0.13 & (0.896) & 0.15 &  (0.880)   \\
&  & RLF &1.46 &  (0.144) & 1.08 &  (0.280) & 1.31  & (0.190) & 1.60  & (0.110)   \\
& & QLIKE &1.72  & (0.085) & 1.54  & (0.123) & 1.64  & (0.101) & 1.82 &  (0.068)  \\
\cline{2-11}
& \multirow{3}{*}{\begin{tabular}{c} Logit\\  $\tau=5$ \end{tabular}} & MSE& 2.38 &  (0.017) & 2.36  & (0.018) & 2.38  & (0.017) & 2.39  & (0.016)   \\
& & RFL & 3.77 & $(<0.001)$  & 3.74 &  $(<0.001)$ & 3.76  & $(<0.001)$ & 3.78  & $(<0.001)$   \\
& & QLIKE&6.01 &  $(<0.001)$ & 5.94 &  $(<0.001)$ & 5.90  & $(<0.001)$ & 5.92  & $(<0.001)$  \\
\cline{2-11}
& \multirow{3}{*}{\begin{tabular}{c}   Logit\\  $\tau=15$ \end{tabular}}  & MSE& 1.45 &  (0.147) & 1.41  & (0.158) & 1.43  & (0.152) & 1.46 &  (0.144)   \\
& & RFL &3.18 &  $(<0.001)$ & 2.99 &  $(0.002)$ & 3.11  & $(0.002)$ & 3.27 &  $(<0.001)$   \\
& & QLIKE & 6.46 &  $(<0.001)$ & 6.14  & $(<0.001)$ & 6.30  & $(<0.001)$ & 6.71 &  $(<0.001)$  \\
\cline{2-11}
& \multirow{3}{*}{\begin{tabular}{c}   Logit\\  $\tau=25$ \end{tabular}}  & MSE& 2.35 &  (0.018) & 2.28  & (0.022) & 2.33  & (0.012) & 2.40 &  (0.016)   \\
& & RFL &5.18 &  $(<0.001)$ & 4.97 &  $(<0.001)$ & 5.09  & $(<0.001)$ & 5.23 &  $(<0.001)$   \\
& & QLIKE & 7.89 &  $(<0.001)$ & 7.52  & $(<0.001)$ & 7.68  & $(<0.001)$ & 7.83 &  $(<0.001)$  \\
\cline{2-11}
& \multirow{3}{*}{\begin{tabular}{c} Logit\\ Combination \end{tabular}}  & MSE& 2.40 &  (0.016) & 2.27  & (0.023) & 2.36  & (0.017) & 2.47 &  (0.013)   \\
& & RFL &4.95 &  $(<0.001)$ & 4.69 &  $(<0.001)$ & 4.84  & $(<0.001)$ & 5.05 &  $(<0.001)$   \\
& & QLIKE & 8.09 &  $(<0.001)$ & 7.78  & $(<0.001)$ & 7.87  & $(<0.001)$ & 8.09 &  $(<0.001)$  \\
\hline
\end{tabular}
\end{adjustbox}
\end{table}
In Table \ref{tunk5}, we extend our analysis by including a Markov-Switching Garch model. The null hypothesis of equal predictability between the logit and the MS-Garch specification is also rejected, indicating a similar pattern as seen in Table \ref{tunk4}, which provides evidence that these models are not equally applicable for volatility forecasting exercises. For the Aranda-Ordaz DDMS specification, our qualitative results also hold when compared to the plain vanilla Garch, with a few exceptions depending on the level of null hypothesis rejection.
\begin{table}[h!]	
\caption{Results from the Diebold-Mariano-West test applied to the MS-Garch model. Presented are the $t$-statistics from Diebold-Mariano-West test along with their $p$-value in parentheses.
\vspace{.5cm}}\label{tunk5}
\centering
\renewcommand{\arraystretch}{1.3}
\begin{adjustbox}{max width=\textwidth}
\begin{tabular}{c| c|c|rc|rc|rc|rc}
\hline
\multicolumn{2}{c|}{Models}& Loss  & \multicolumn{2}{c|}{MedRV}  & \multicolumn{2}{c|}{MinRV}  & \multicolumn{2}{c|}{BV}  & \multicolumn{2}{c}{RV}     \\
\hline
\multirow{15}{*}{MS-Garch} & \multirow{3}{*}{\begin{tabular}{c} A-O\\  $\tau=5$ \end{tabular}} & MSE & 0.62 &  (0.535) & 0.08  & (0.936)  & 0.40 & (0.689) & 0.90 &  (0.368)   \\
&  & RLF &2.10 &  (0.036) & 1.65 &  (0.098) & 1.94  & (0.052) & 2.26  & (0.024)   \\
& & QLIKE &1.72  & (0.085) & 1.54  & (0.124) & 1.69  & (0.091) & 1.98 &  (0.048)  \\
\cline{2-11}
& \multirow{3}{*}{\begin{tabular}{c} Logit\\  $\tau=5$ \end{tabular}} & MSE& 2.43 &  (0.015) & 2.40  & (0.016) & 2.43  & (0.015) & 2.44  & (0.015)   \\
& & RFL & 3.99 & $(<0.001)$  & 3.97 &  $(<0.001)$ & 3.99  & $(<0.001)$ & 3.99  & $(<0.001)$   \\
& & QLIKE&6.14 &  $(<0.001)$ & 6.05 &  $(<0.001)$ & 6.06  & $(<0.001)$ & 6.15  & $(<0.001)$  \\
\cline{2-11}
& \multirow{3}{*}{\begin{tabular}{c}   Logit\\  $\tau=15$ \end{tabular}}  & MSE& 1.49 &  (0.136) & 1.45  & (0.147) & 1.48  & (0.139) & 1.50 &  (0.134)   \\
& & RFL &3.35 &  $(<0.001)$ & 3.18 &  $(0.002)$ & 3.30  & $(0.001)$ & 3.42 &  $(<0.001)$   \\
& & QLIKE & 6.45 &  $(<0.001)$ & 6.15  & $(<0.001)$ & 6.28  & $(<0.001)$ & 6.63 &  $(<0.001)$  \\
\cline{2-11}
& \multirow{3}{*}{\begin{tabular}{c}   Logit\\  $\tau=25$ \end{tabular}}  & MSE& 2.41 &  (0.016) & 2.34  & (0.019) & 2.39  & (0.017) & 2.47 &  (0.013)   \\
& & RFL &5.30 &  $(<0.001)$ & 5.12 &  $(<0.001)$ & 5.23  & $(<0.001)$ & 5.34 &  $(<0.001)$   \\
& & QLIKE & 7.86 &  $(<0.001)$ & 7.49  & $(<0.001)$ & 7.67  & $(<0.001)$ & 7.91 &  $(<0.001)$  \\
\cline{2-11}
& \multirow{3}{*}{\begin{tabular}{c} Logit\\ Combination \end{tabular}}  & MSE& 2.54 &  (0.011) & 2.40  & (0.016) & 2.51  & (0.012) & 2.61 &  (0.009)   \\
& & RFL &5.35 &  $(<0.001)$ & 5.09 &  $(<0.001)$ & 5.25  & $(<0.001)$ & 5.42 &  $(<0.001)$   \\
& & QLIKE & 8.08 &  $(<0.001)$ & 7.74  & $(<0.001)$ & 7.85  & $(<0.001)$ & 8.16 &  $(<0.001)$  \\
\hline
\end{tabular}
\end{adjustbox}
\end{table}
\subsubsection*{Dynamic Predictive Accuracy}
The forecasting period exhibits notable intra-state fluctuations, including bear rallies (positive sub-trends) and bull corrections (negative sub-trends), alongside high uncertainty and volatility. However, the previous analysis based on the Diebold-Mariano-West test only considers the entire sample, neglecting possible intra-period uncertainties. Given the nature of the forecasting period, it is interesting to assess the evolution of the relative forecasting performance between the proposed DDMS model and alternative Garch specifications over time. Thus, we conduct the \citet{giacomini_rossi} fluctuation test to compare this temporal evolution. For point forecasts, this approach involves a rolling version of the Diebold-Mariano-West test, computed with a rolling window representing 10\% of the total out-of-sample sample size.

Figure \ref{fig_fluc1} shows the evolution of the test statistic for the null hypothesis of equal predictability between the Aranda-Ordaz DDMS model and competing Garch models for three considered loss functions and using the RV proxy.\footnote{The test results are nearly identical when different proxies are used.}  When the test statistic exceeds the positive critical value, it offers evidence in favor of the Aranda-Ordaz model; when negative, it favors the Garch models.
\begin{figure}[h!]	
	\centering
\mbox{
	\subfigure[MSE]{\includegraphics[width=.3\textwidth]{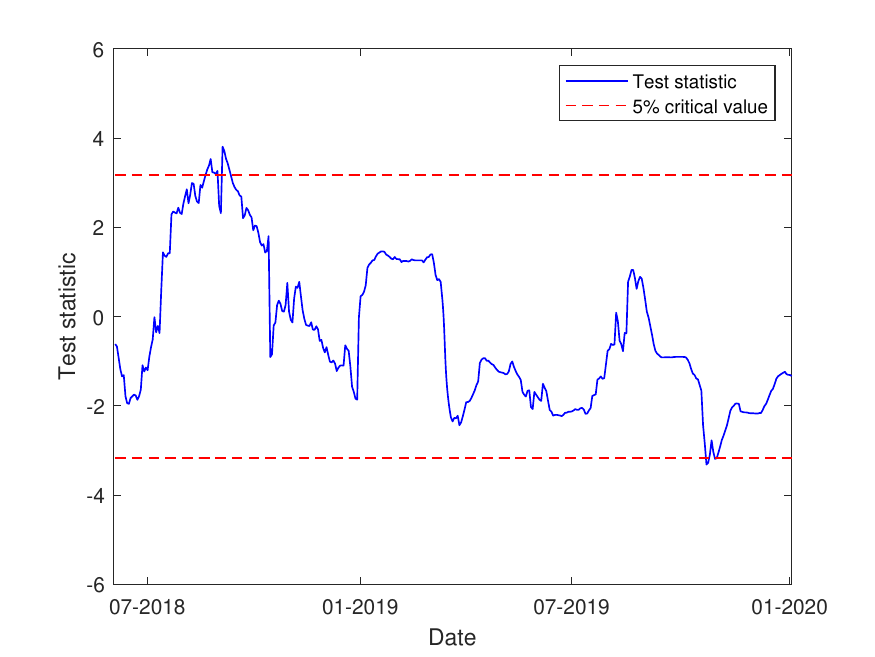}}
	\subfigure[RLF]{\includegraphics[width=.3\textwidth]{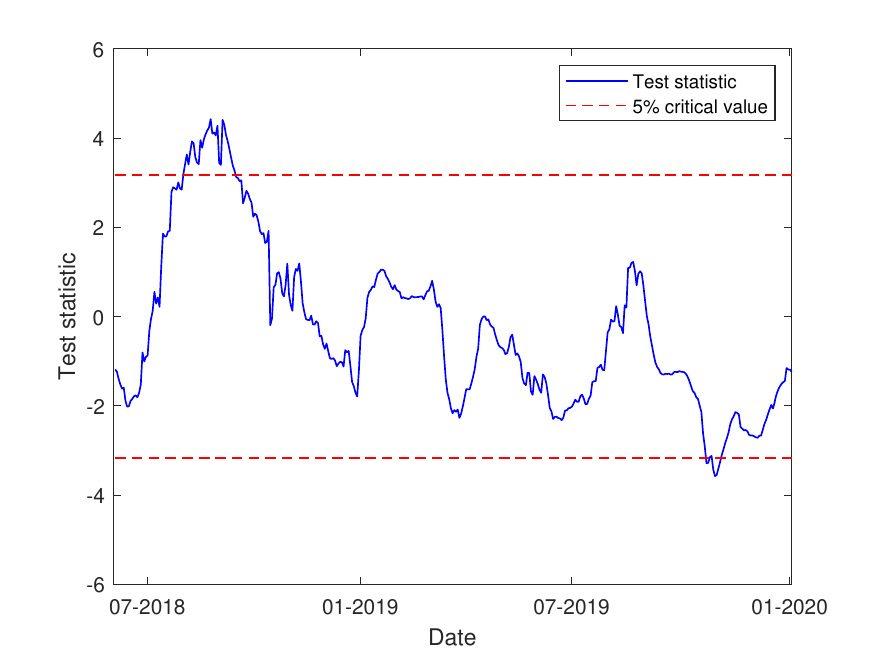}}
	\subfigure[QLIKE]{\includegraphics[width=.3\textwidth]{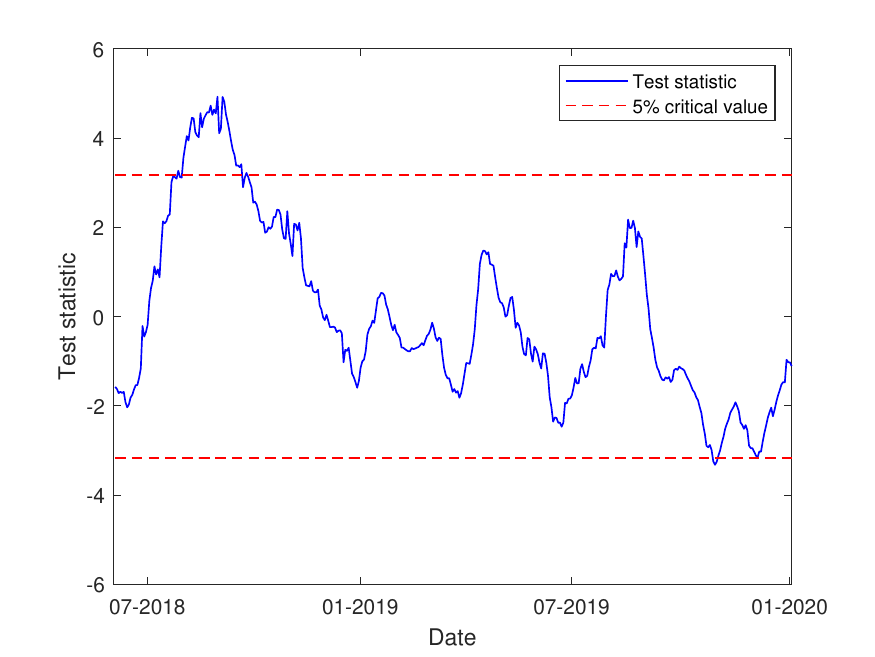}}
}
\mbox{
	\subfigure[MSE]{\includegraphics[width=.3\textwidth]{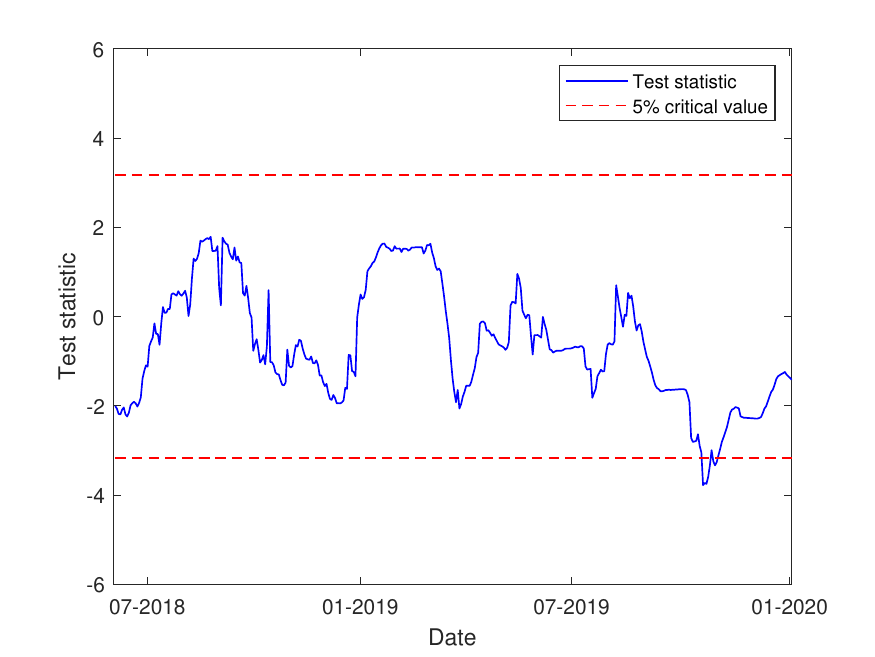}}
	\subfigure[RLF]{\includegraphics[width=.3\textwidth]{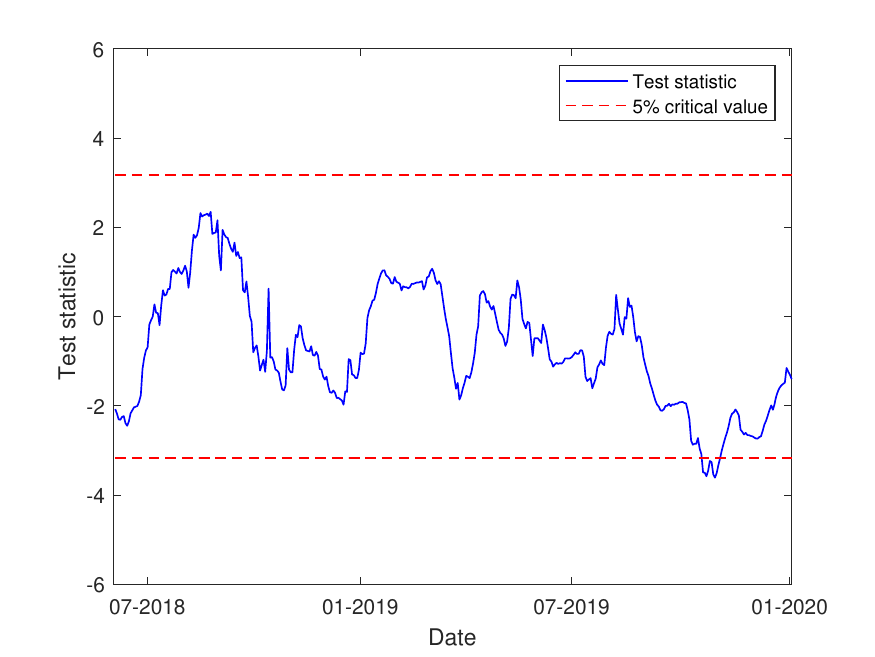}}
	\subfigure[QLIKE]{\includegraphics[width=.3\textwidth]{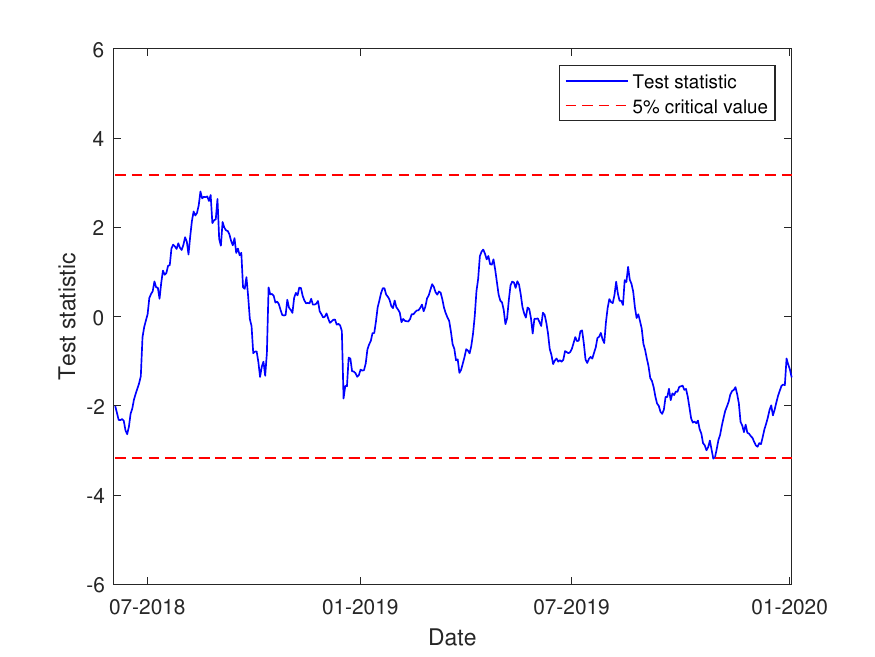}}
}
	\caption{Fluctuation test between the DDMS A-O and Garch  (top row) and DDMS A-O and MS-Garch (bottom) for different loss functions. }	 \label{fig_fluc1}
\end{figure}

In summary, the Aranda-Ordaz DDMS model shows superior performance over the standard Garch model in the initial phase of the forecasting sample, particularly for the QLIKE loss function, where it maintains an advantage for approximately 40 days. For most of the sample period, both models exhibit similar predictive accuracy. On the other hand, the Garch model outperforms the Aranda-Ordaz DDMS model in two occasions in 2019. The comparison with the MS-Garch model also highlights that, for most of the sample size, both models have similar forecasting accuracy. These results suggest that the DDMS model with the Aranda-Ordaz link function is a competitive alternative to Garch models, with the potential to outperform them during certain forecasting periods.
\section{Conclusion}
This paper proposes a methodology to address the issue of duration choice in DDMS models. The methodology involves the application of a parametric link function in place of the typical fixed link function to calculate transition probabilities. The proposed model present higher likelihood values and is capable of a more accurate description of transition probabilities. The proposed approach is capable of significantly improving forecasting accuracy, especially in cases of duration misspecification.

Two Monte Carlo simulations, based on classical applications of DDMS models, are presented. The results suggests that using the Aranda-Ordaz link function leads to more precise forecasts and transition probabilities, not only in cases of duration misspecification but also when the model is correctly specified. Furthermore, an empirical study is conducted to forecast the volatility of the S\&P500, which illustrates the effectiveness of the proposed methodology.

The empirical results indicate that the Aranda-Ordaz DDMS model outperforms the traditional fixed logit transitions in terms of forecast precision for most duration parameters. Moreover, the Aranda-Ordaz link function improves the forecasting performance to such an extent that it is equivalent to the Garch-type models. Notably, such an improvement in forecasting accuracy is not observed when using fixed link functions, revealing that the methodology boosts the duration-dependent specifications in this direction.

Besides our modeling approach, we have not observed any empirical research that investigates the predictive ability of DDMS models in a point volatility forecasting exercise, as we conducted. However, different extensions of this paper can be pursued. This includes assessing the performance of the Aranda-Ordaz DDMS in other economic applications, such as risk assessment through Value-at-Risk (VaR) estimation or constructing investment strategies based on the regime probabilities of these classes of models, as seen in the trading strategies literature.

\bibliographystyle{apalike}
\bibliography{mybibfilef}

\section*{Appendix: Maximum Likelihood Estimation of DDMS Models}

This appendix briefly describes the estimation procedure of the DDMS model as outlined in \cite{maheu_identifying_2000}. Consider the following specification for the DDMS model
\begin{equation}\label{gt}
{Y_t} = {\mu _0}(1 - {S_t}) + {\mu _1}{S_t} + \sigma {Z_t},\quad {Z_t} \sim N(0,1),
\end{equation}
where $\sigma$ is the standard deviation parameter and  ${Z_t}$ is assumed to follow an identically and independently normal distribution.
Observe that in \eqref{gt} we have $\mu(0)=\mu_0$ and $\mu(1)=\mu_1$. The DDMS model can be viewed as an extension of Hamilton's model \citep{hamilton_new_1989}, as a new latent variable, $\mathcal{S}_t$, encompasses all possibilities of historical trajectories from $S_t$ to $\tau$:
\begin{align*}
\mathcal{S}_t=1,\quad &\mbox{ if }\quad S_{ t }=1,S_{ t-1 }=0,S_{t-2},\cdots,D(S_{ t })=1, \\
\mathcal{S}_t=2,\quad &\mbox{ if }\quad S_{ t }=1,S_{ t-1 }=1,S_{t-2},\cdots,D(S_{ t })=2, \\
\mathcal{S}_t=3,\quad &\mbox{ if }\quad S_{ t }=0,S_{ t-1 }=1,S_{t-2}, \cdots, D(S_{ t })=1,\\
&\ \ \vdots \\
\mathcal{S}_t=N,\quad &\mbox{ if }\quad S_{ t }=0,S_{ t-1 }=0,S_{t-2}=0, \cdots,D(S_{ t })=\tau,
\end{align*}
where $N := 2 + 2(\tau -  1)$ are the extended states. Let $\bs\xi_t := \big(I(\mathcal{S}_t=1), \cdots, I(\mathcal{S}_t=N)\big)'$, where $I$ denotes the indicator function. Let $\mathcal P$ be the $N\times N$ transition matrix associated to $\mathcal S_t$, whose $(i,j)$-$th$ entry is given by $[P]_{i,j}= P(\mathcal{S}_t = j | \mathcal{S}_{t-1} = i)$. Let $\F _t:=\sigma(Y_t, Y_{t-1}, \cdots, S_t,S_{t-1},\cdots)$ denote the information available at time $t$ and let  $f(Y_t|\mathcal{S}_t = j, \F _{t-1})$ be the conditional density of $Y_t$.  Let
\begin{equation*}
\bs\eta_t:=\big(f(Y_t | \mathcal{S}_t = 1, \F _{t-1}),\cdots,f(Y_t | \mathcal{S}_t = N, \F _{t-1})\big)',
\end{equation*}
and $\hat{\bs\xi}_{s|r}$ be a vector for which the $j$-th element is $P(\mathcal{S}_s = j | \F _{r})$.  Observe that we have
\begin{align*}
\hat{\bs\xi}_{t|t} =\frac{\hat{\bs\xi}_{t|t-1} \odot \bs\eta_t}{\bs1'_N(\hat{\bs\xi}_{t|t-1} \odot \bs\eta_t)}, \qquad \mbox{and}\qquad \hat{\bs\xi}_{t+1|t} & =  \mathcal P \hat{\bs\xi}_{t|t},
\end{align*}
where $\bs 1_N:=(1,\cdots,1)'\in\R^N$. Following \cite{ham}, we start with
\begin{equation*}
{\bs\xi}_{0|0}= (A'A)^{-1} A'E,
\end{equation*}
where $A=\begin{bmatrix} I_{N}-{\mathcal P}\\ \bs 1_N' \end{bmatrix}$ e $E=\begin{bmatrix} \bs 0_N\\ 1 \end{bmatrix}$, with $\bs0_N:=(0,\cdots,0)'\in\R^N$ and $I_N$ denoting the $N\times N$ identity matrix. Finally, the conditional log-likelihood function is given by
\begin{equation*}
L(\theta) = \sum_{t=1}^n \log\big( f(Y_t |\F _{t-1})\big),
\end{equation*}
where
\begin{equation*}
f(Y_t |\F _{t-1}) = \bs1_N'(\hat{\bs\xi}_{t|t-1} \odot \bs\eta_t).
\end{equation*}


\end{document}